\documentclass[aps,pra,showpacs,notitlepage,onecolumn,superscriptaddress,nofootinbib]{revtex4-1}
\usepackage[utf8]{inputenc}
\usepackage[tmargin=1in, bmargin=1in, lmargin=1.25in, rmargin=1.25in]{geometry}
\usepackage{amsmath, amssymb, amsthm}
\usepackage{graphicx}
\usepackage{xcolor}
\usepackage{enumitem}
\usepackage{titlesec}
\usepackage{datetime}
\usepackage{hyperref}
\usepackage{import}
\usepackage{mathtools}

\definecolor{crimson}{RGB}{186,0,44}

\hypersetup{
    colorlinks,
    linkcolor={crimson},
    citecolor={crimson},
    urlcolor={crimson}
}

\theoremstyle{definition}
\newtheorem{definition}{Definition}[section]
\newtheorem{lemma}{Lemma}[section]
\newtheorem{theorem}{Theorem}[section]

\newtheorem*{theorem*}{Theorem}
\newtheorem*{corollary*}{Corollary}
\newtheorem{remark}{Remark}[section]

\newtheorem{problem}{Problem}[section]

\begin{document}

\title{Quantum signal processing with continuous variables}
\author{Zane M.\ Rossi}
\affiliation{%
Department of Physics, Massachusetts Institute of Technology, Cambridge, Massachusetts 02139, USA}
\affiliation{%
Physics and Informatics Laboratory, \mbox{NTT Research,~Inc.,}
940 Stewart Dr., Sunnyvale, California, 94085, USA}
\affiliation{%
NTT Basic Research Laboratories and Research Center for Theoretical Quantum Physics,
3-1 Morinosato-Wakamiya, Atsugi, Kanagawa 243-0198, Japan}
\author{Victor M.\ Bastidas}\affiliation{%
NTT Basic Research Laboratories and Research Center for Theoretical Quantum Physics,
3-1 Morinosato-Wakamiya, Atsugi, Kanagawa 243-0198, Japan}
\author{William J.\ Munro}\affiliation{%
NTT Basic Research Laboratories and Research Center for Theoretical Quantum Physics,
3-1 Morinosato-Wakamiya, Atsugi, Kanagawa 243-0198, Japan}
\author{Isaac L.\ Chuang}\affiliation{%
Department of Physics, Massachusetts Institute of Technology, Cambridge, Massachusetts 02139, USA}
\date{\today}

\begin{abstract}
    \noindent Quantum singular value transformation (QSVT) enables the application of polynomial functions to the singular values of near arbitrary linear operators embedded in unitary transforms, and has been used to unify, simplify, and improve most quantum algorithms. QSVT depends on precise results in representation theory, with the desired polynomial functions acting simultaneously within invariant two-dimensional subspaces of a larger Hilbert space. These two-dimensional transformations are largely determined by the related theory of quantum signal processing (QSP). While QSP appears to rely on properties specific to the compact Lie group SU(2), many other Lie groups appear naturally in physical systems relevant to quantum information. This work considers settings in which SU(1,1) describes system dynamics and finds that, surprisingly, despite the non-compactness of SU(1,1), one can recover a QSP-type ansatz, and show its ability to approximate near arbitrary polynomial transformations. We discuss various experimental uses of this construction, as well as prospects for expanded relevance of QSP-like ansätze to other Lie groups.
\end{abstract}

\maketitle

\section{Introduction}

\noindent In quantum computing the Lie group associated with the evolution of a single qubit, SU(2), has received the majority of attention and study—this group and its related algebra permit basic intuition for the character of certain quantum computations, mainly through the ubiquitously applied surjective homomorphism from SU(2) to SO(3), diagrammed as the Bloch-sphere. In some quantum algorithms, e.g., Grover search \cite{grover_05, hoyer_00}, the evolution of a multiple-qubit systems can be simplified to two-dimensional transformations, by which the pleasant properties of SU(2) are recovered. The algorithmic techniques at the center of this work, quantum signal processing (QSP) \cite{lyc_16_equiangular_gates, lc_17_simultation, lc_19_qubitization} and its lifted version quantum singular value transformation (QSVT) \cite{gslw_19}, permit similar SU(2)-derived intuition.

QSP and QSVT have seen success in unifying, simplifying, and improving most known quantum algorithms \cite{mrtc_21}, in turn showcasing that basic properties of SU(2), when properly understood and applied, are sufficient to capture unexpectedly sophisticated algorithmic behavior. These algorithms, by use of a simple alternating circuit ansatz, permit one to modify the singular values of near arbitrary linear operators by polynomial functions; while abstract, this fundamental linear algebraic manipulation subsumes algorithms for Hamiltonian simulation \cite{coherent_ham_sim_21}, phase estimation \cite{rall_21}, quantum-inspired machine learning algorithms \cite{chia_20}, semi-definite programming \cite{q_sdp_solvers_20}, adiabatic methods \cite{lin_eig_filter_20}, computation of approximate correlation functions \cite{rall_correlation_20}, computation of approximate fidelity \cite{gilyen_fidelity_22}, recovery maps \cite{petz_recovery_20}, metrology \cite{dgn_qsp_metrology_22}, and fast inversion of linear systems \cite{tong_inversion_21}.

The theory of QSP has roots in the study of composite pulse techniques for NMR \cite{wimperis_bb1_94, ylc_14, lyc_16_equiangular_gates} but was first named for use in Hamiltonian simulation \cite{lc_17_simultation, lc_19_qubitization}, in which the single-qubit nature of QSP was suitably lifted to apply to systems of multiple qubits by a technique known as qubitization. This idea was greatly expanded to cover the manipulation of general, non-normal linear operators and termed QSVT \cite{gslw_19}, by which one can achieve QSP-like manipulation of invariant SU(2) subspaces preserved by alternating projectors according to Jordan's lemma \cite{jordan_75}. Recently this argument has been even further simplified in relation to the cosine-sine decomposition \cite{cs_qsvt_tang_tian}.

Parallel to this development, experimental work in quantum optics has long considered basic interferometric operations, whose action on optical modes are also describable by SU(2). In modifying these passive interferometric devices to actively driven ones, one can move from an SU(2) description to one desfined by the related but non-compact Lie group SU(1,1). Such devices have the upshot of enabling improved sensitivity for a variety of interferometric measurements, as well as greatly simplified experimental apparatuses \cite{yurke_su2_su11_86}. This apparently simple change in the defining algebra has deep experimental and theoretical implications, and correspondingly the general analysis of composite systems of SU(1,1) transformations is difficult \cite{ou_li_su11_review_20, su_mode_engineering_19} beyond low-gain regimes.

In light of this, it is worthwhile to determine whether QSP and QSVT, which both (1) simply subsume a large number of quantum algorithmic techniques and (2) rely strongly on basic properties of SU(2), can be suitably modified to apply to and analogously simplify the analysis composite systems of similar SU(1,1) interactions. While this extension may appear mathematically mild, the loss of the compactness of SU(2), as well as movement from finite-dimensional unitary operations to those involving continuous variables, is significant. In bridging the gap from SU(2) to SU(1,1) for QSP-like algorithms, this work gives first steps in addressing the significant theoretical challenges and correspondingly curious insights within the application of QSP and QSVT to continuous variable quantum computation.

This work is somewhat intended for an audience already familiar with the basic structure and theorems of standard QSP and QSVT. However, we reproduce some salient theorems from previous works in Appendix~\ref{appendix:basics_qsp_qsvt}, and discuss and reproduce proof techniques where appropriate. As this work concerns a modification to some of the low-level tenets of these algorithms, we try to phrase our constructions in terms of what they preserve from previous work, and what they break (and thus force us to recover and re-prove). In many senses the QSP and QSVT ansätze are fragile, and modifying them to describe new contexts introduces subtleties, various no-go results, and a few exciting insights.

\begin{figure}[h!]
    \centering
    \includegraphics[width=0.75\textwidth]{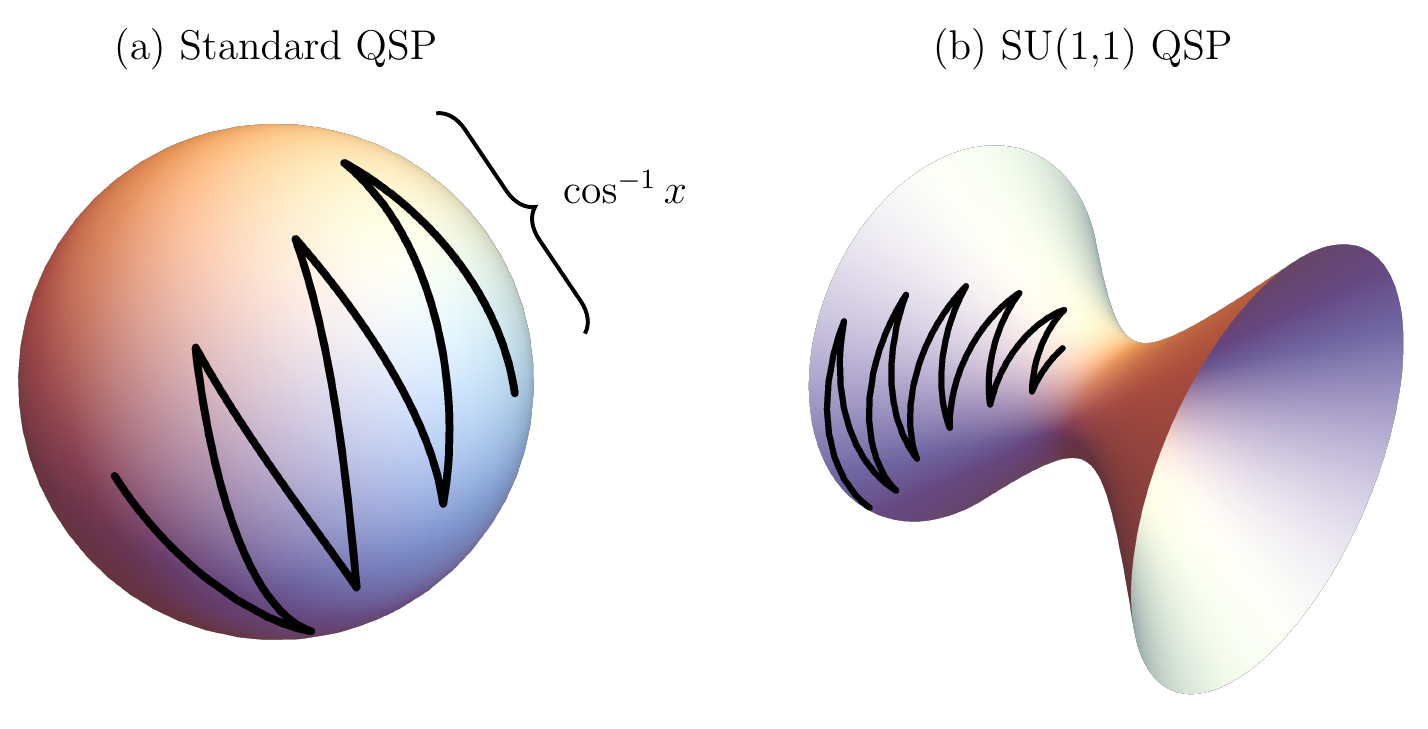}
    \caption{Conceptual depictions of the actions of (a) standard QSP and (b) SU(1,1) QSP according to the natural surjections (a) SU(2) to SO(3) and (b) SU(1,1) to SO(2, 1). I.e., elements of each group can be recast as preserving the unit sphere or an (infinitely extending) hyperbola respectively. QSP, for a given argument $x$, is a walk on the Bloch sphere with fixed step angle $\theta = \cos^{-1}{x}$ and step direction $\phi_k$ the $k$-th QSP phase. For SU(1,1) QSP, this walk can be seen on a hyperbola, with generalized step angle $\beta = \cosh^{-1}{x}$; that the manifold for (b) is not compact is a major source of the difference in the corresponding theories, and the reason for the modified ranges of the maps from $x$ to $\theta, \beta$. In both cases walk steps are along geodesics.}
    \label{fig:qsp_as_walk}
\end{figure}

\begin{figure}
    \centering
    \includegraphics[width=1.0\textwidth]{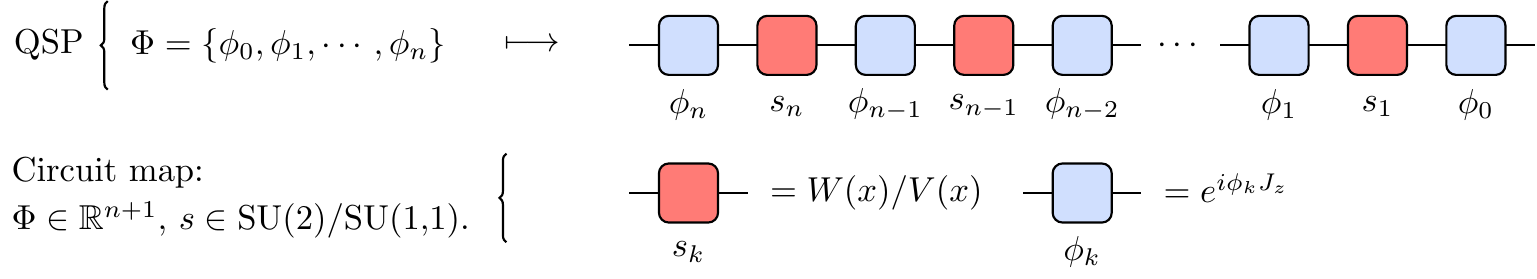}
    \caption{The abstract form of a QSP protocol. As discussed in Fig.~\ref{fig:qsp_as_walk}, both standard QSP and its SU(1,1) variant can be seen as a walk on a manifold. Namely, one interleaves controllable (blue, $\phi_k$) elements of a Lie group and an unknown but consistent (red, $s_k$) oracle operation, such that the ultimate unitary depends strongly on the unknown signal $s$. Here $s$ relates simply to the phased iterates $W_\phi(\theta)$ and $V_\phi(\beta)$ up to $\theta = \cos^{-1}(x)$ and $\beta = \cosh^{-1}(x)$, where in Eq.~\ref{eq:phased_iterates} abutting phases have been absorbed into the signal.}
    \label{fig:qsp_generalized_circuit}
\end{figure}

QSP relies strongly on familiar aspects of SU(2); its comprising gates are rotations on the Bloch sphere, and its action is, despite application to multiple-qubit settings, summarized by the evolution of a single qubit. For the remainder of this work we take $X, Z$ as the Pauli matrices (one representation of the generators of $\mathfrak{su}(2)$) with the common form
    \begin{equation}
        X = 
        \begin{bmatrix}
            0 & 1\\
            1 & 0
        \end{bmatrix},
        \quad
        Y = 
        \begin{bmatrix}
            0 & -i\\
            i & 0
        \end{bmatrix},
        \quad
        Z = 
        \begin{bmatrix}
            1 & 0\\
            0 & -1
        \end{bmatrix}.
    \end{equation}
QSP protocols can be visualized as walks on the Bloch sphere, where the distance per step is defined by the unknown signal $x = \cos{\theta}$, and the direction of each step by the known and chosen $\phi_k \in \mathbb{R}$ for $k \in [n]$ where $n \in \mathbb{N}$ is the length of the QSP protocol. The power of the theory of QSP lies in that the action of this walk can be precisely controlled such that the same choice of direction for each step results, when the step length is changed, in drastically different behavior. This is depicted in Fig.~\ref{fig:qsp_as_walk}, with more explicit description of these protocols in Fig.~\ref{fig:qsp_generalized_circuit}.

The surprising aspect of the theory of QSP is that one can quickly and classically compute the proper $\phi_k$ such that, for walks of a given step size $x$, said \emph{fixed set of $\phi_k$} achieve a near arbitrary desired end point for said walk. In other words, QSP permits immense control over functions of the form
    \begin{equation}
        \text{QSP} : \mathbb{R}^{n+1} \,\rightarrow\, (F : [-1, 1] \,\rightarrow\, \text{SU(2)}).
    \end{equation}
Here we mean that QSP is a map from an ordered list of real numbers (the QSP phases $\Phi$) to a function $F$ from the interval to the Lie group SU(2). This basic type signature exemplifies QSP's utility as a generator of \emph{superoperators}. The ansatz takes in a set of parameters and returns a function taking a scalar (encoded as a rotation) to an element of the compact Lie group SU(2) in a tunable way. In fact, one can prove the following result summarizing the expressivity of the QSP ansatz.
    \begin{theorem}[Expressivity of the QSP ansatz] \label{thm:qsp_ansatz_expressivity}
        The set of functions achieved by the set of QSP protocols of finite length is dense in the set of definite-parity piecewise-continuous functions with the form $[-1, 1] \rightarrow SU(2)$, up to an ambiguity (a rotation about a known, fixed axis on the Bloch sphere) parameterized by a single function with the form $R: [-1, 1] \rightarrow U(1)$. Additionally, the rate of uniform convergence to a desired function $G: [-1, 1] \rightarrow SU(2)$ up to this ambiguity is inverse polynomial in protocol length in the worst case, and inverse exponential under certain common assumptions of smoothness of the desired functional form.
    \end{theorem}
We claim that the path from QSP to its lifted counterpart, quantum singular value transformation (QSVT) is simple enough to preserve much of the single-qubit character of standard QSP. In QSVT, the geometric intuition of QSP is preserved and lifted almost solely due to a classic result of Jordan \cite{jordan_75} (with modern proofs in \cite{regev_06, cs_qsvt_tang_tian}). This lemma states that products of two reflections necessarily preserve one- and two-dimensional subspaces, which in the context of QSVT means that circuits can be designed which implicitly perform QSP-like SU(2) unitary operations within such subspaces. These unitary operations have a natural interpretation as modifying the singular values of block-encoded linear operators by precisely the same polynomial functions achievable with standard QSP. In this case, the encoded linear operators can be thought of as mappings between sets of left and right singular vectors, where one is assumed to have easy access to projectors onto the \emph{span} of these two sets respectively. 

While QSVT has enabled a qualified unification of quantum algorithms \cite{mrtc_21}, it could also be seen as indicating a scarcity of diverse quantum algorithmic techniques. Consequently, probing related results in functional analysis and representation theory has potential, guided by the success of QSVT, to generate novel quantum algorithms. Indeed, this work shows that even basic modifications to the QSP ansatz can induce large changes in the character of the resulting algorithm. Here the change specifically concerns the natural Lie group for elements of the ansatz.

We can now pose an informal problem statement for this work, relating to the two Lie groups SU(2) and SU(1,1). It is known that quantum signal processing (QSP) and quantum singular value transformation (QSVT) rely strongly on properties of SU(2), the former consisting entirely of interleaved products of elements of this group. Moreover, SU(2) and SU(1,1) are known to be quite similar algebraically (in terms of their generators) despite appearing in substantively different physical contexts, and the latter being non-compact. Together, these two observations point toward the following problem statement.

\begin{problem}[Informal problem statement]
    Do techniques similar to those used in the theory of QSP permit one to \emph{usefully characterize} interleaved products of elements of SU(1,1)?
\end{problem}

While we claim an answer of `yes' to this problem statement, how we have stated it brings up a few important ambiguities. The first is that we desire a \emph{useful characterization} of such products. In standard QSP this characterization manifests in showing which polynomials in an unknown parameter for an oracle unitary can be embedded as matrix elements. In our setting we slightly modify this condition, asking not only about the description of analogous polynomials, but their density other functional spaces, as well as use in approximating arbitrary desired functions. Moreover, as SU(1,1) lacks finite dimensional unitary representations, we will have to do some lifting in resituating what we mean by embedding a polynomial transform; this is the purview of Sec.~\ref{sec:analyzing_qsp_ansatz}. Secondly, in this new setting we may not be bound to interleaved products of the same form as in QSP, or with the same physical interpretation; defining our ansatz, and discussing its physical reasonableness, is the purview of Sec.~\ref{sec:qsp_ansatz}. Together, these two sections give a concrete, formal setup and analysis of this movement of the theory of QSP from SU(2) to SU(1,1).

We stress that modifying QSP to encompass SU(1,1) dynamics is not a superficial translation; while the linear operators describing such interleaved products may appear similar, the non-compactness of SU(1,1) puts into question a vital but overlooked result in standard QSP. I.e., most results in the theory of functional approximation and its famous cousin Fourier analysis rely on natural bases of functions which are either (1) periodic or (2) are square integrable. Critically, the SU(1,1) analogue of standard QSP sacrifices both of these traits along with its compactness, for fundamental reasons in the theory of Lie groups \cite{wigner_39}. Thus to recover a useful theory of SU(1,1) QSP, we need to provide a bridge from the achievable polynomials with this ansatz to the approximable functions. Spanning this gap (which did not exist in standard QSP) is the heart of this paper, and the focus of Sec.~\ref{sec:analyzing_qsp_ansatz}.

\subsection{Prior work} \label{sec:prior_work}

Given the current understanding QSVT as a quantum algorithm, room for research has fallen along two main axes. Along the first, one can translate classical or quantum algorithmic tasks to the language of QSVT, and prove or disprove (possibility with respect to careful assumptions) the existence of efficient block-encodings (ways to load linear operators into the QSVT ansatz for processing) and polynomial transforms required to solve given algorithmic tasks. This axis has experienced great success \cite{coherent_ham_sim_21, rall_21, lombardi_pqzk_2021, chia_20, q_sdp_solvers_20, lin_eig_filter_20, rall_correlation_20, gilyen_fidelity_22, petz_recovery_20, dgn_qsp_metrology_22, tong_inversion_21} and enabled improvements in query complexity lower bounds for a variety of famed quantum algorithms. The second axis involves taking the basic mathematical tenants which enabled the success of the theory of QSVT, and \emph{augmenting or deepening them} to apply to contexts not previously amenable to QSVT. While this latter work remains more exploratory \cite{rc_22, sym_qsp_21, dlnw_infinite_22, tltc_alec_23, rycs_noisy_22}, it has demonstrated significant creative potential for QSP- and QSVT-inspired methods for substantively different circuit ansätze. One such unexplored generalization concerns transporting QSP and QSVT to algebraically distinct settings.

To map algorithmic problems to functional analytic ones, QSP relies strongly on special properties of SU(2), most notably the simple relations between its generators, and the compactness of the Lie group. What remains unclear, however, is whether this map can survive the jump to more exotic (but still physically reasonable) Lie group and algebras. While not extensively discussed in this paper, the guiding motivation for this work rests on the dual appearance of both SU(2) and SU(1,1) in photonic systems \cite{yurke_su2_su11_86, ou_li_su11_review_20}. There exist multiple works examining the serial application of elements from both of these Lie groups, showing, albeit only non-analytically or perturbatively, that such protocols permit the useful manipulation of continuous variable quantum information \cite{su_mode_engineering_19, cui_mode_engineering_20}. This work re-examines these settings, showing that many of the strong statements possible in QSP due to the simplicity of the underlying Lie algebra, can be ported to similar (but notably non-compact) algebras, and that moreover the resulting protocols have reasonable physical interpretation and a concise analytic characterization. As far as the authors are aware, this is the first work describing such cascaded SU(1,1) interactions in the large-amplification regime, as well as the first application of QSP-like methods to non-compact Lie groups.

\subsection{Outline of results} \label{sec:summary_of_results}

This work is broken into three major sections. The first, in Sec.~\ref{sec:qsp_ansatz}, discusses the explicit form of an SU(1,1) analogue to QSP, providing a series of supporting definitions and lemmas from the theory of Lie groups and algebras. The second major portion discusses the expressive power of this ansatz: in other words, Sec.~\ref{sec:analyzing_qsp_ansatz} discusses the ability of this ansatz to embed polynomial transforms of its matrix elements. This section calls on further results in the theory of Lie groups and generalized Fourier analysis, and provides compact proofs where possible, toward an explicit description of the family of achievable functional transforms. The flavor of the results of this work is summarized by the following informal statements:
    \begin{enumerate}[label=(\arabic*)]
        \item The achievable polynomials in SU(1,1) QSP are dense in the space of continuous functions with (a) definite parity and (b) bounded from below by one.
        \item However, the length of an SU(1,1) QSP protocol which uniformly approximates a desired continuous function on a \emph{finite} interval in general must grow exponentially in the size of the interval of uniform approximation.
        \item There exist simple, concrete parameterizations of SU(1,1) QSP of length $n$ which achieve, for signals in a finite interval independent of $n$, polynomial transforms whose magnitude is bounded above by a function independent of $n$.
    \end{enumerate}
The above statements are necessarily qualified by specific statements on the domain and form of the considered functions, which are different than those of standard QSP, and for which we refer the interested reader to the main text. Thirdly, as mentioned, we look at three explicit examples for parameterizations for the SU(1,1) QSP ansatz in Sec.~\ref{sec:worked_numerical_examples}, including those which achieve Chebyshev polynomials, monomials, and generalized bandpass functions. We depict and analytically investigate the behavior of these protocols.

Finally in Sec.~\ref{sec:discussion_conclusion}, we discuss the outlook for SU(1,1) QSP, its limitations, and paths toward the extension of techniques in QSP to further settings in continuous variable quantum computation. Algebraically involved proofs and auxiliary but important results in the theory of Lie groups and generalized Fourier series are relegated to the appendices.

\section{Constructing the SU(1,1) QSP ansatz} \label{sec:qsp_ansatz}

This section serves two purposes. The first is to provide a brief and self-contained introduction to relevant concepts in Lie groups and algebras (any involved proofs again in the appendices), limited to discussion and manipulation of the properties of SU(2) and SU(1,1) specifically. We also use this section, following the brief review of algebraic concepts, to introduce the main problem statement of this work: an SU(1,1) analogue of standard QSP. We use this ansatz to more concretely discuss how these Lie groups can appear in the context of quantum information, providing a simple example to motivate the proposed modified QSP ansatz.

\subsection{Lie groups and algebras}

We give minimal presentations of SU(2) and SU(1,1) as Lie groups, as well as their common definitions in terms of Lie algebras. We discuss common properties of these groups toward their use in the quantum computing protocols, and use these connections to present a clear analogy to the theory of QSP before discussing how differences between constructions of QSP relying on SU(2) and SU(1,1) relate to differences in physical implementation. Unless otherwise noted, common definitions and statements can be found in any introductory textbook on compact Lie groups \cite{fegan_lie_groups_1991}.

\begin{definition}[Lie group] \label{def:lie_group}
    Lie groups are both groups and differentiable manifolds. That is, they locally resemble Euclidean space, and multiplication by group elements and inverses is smooth. In other words, the required binary multiplication operation $\mu$ for a Lie group $G$ is such that
        \begin{equation}
            \mu :: G\times G \mapsto G, \quad \mu(g, h) = gh,
        \end{equation}
    is a smooth mapping of the product manifold $G\times G$ to $G$. With this comes the assumption of a well-defined derivative and the weaker condition of continuity.
\end{definition}

\begin{definition}[Lie algebra] \label{def:lie_algebra}
    Lie groups give rise to Lie algebras, whose formal definition is in terms of the tangent space at the identity of the Lie group. Correspondingly, finite-dimensional Lie algebras correspond to connected Lie groups uniquely up to finite covering (if we choose for the Lie group to be simply connected, this correspondence is indeed unique). Formally Lie algebras are vector spaces $\mathfrak{g}$ over a field $F$ with a bilinear alternating map $[\ast, \ast]$ (the Lie bracket) that satisfies the Jacobi identity.
        \begin{align}
            &[\ast, \ast] :: \mathfrak{g}\times\mathfrak{g} \mapsto \mathfrak{g},\\
            &[x, [y, z]] + [y, [z, x]] + [z, [x, y]] = 0.
        \end{align}
\end{definition}

For our purposes we consider finite dimensional Lie algebras whose Lie bracket is the commutator, and whose representations as linear operators are particularly simple. By the uniqueness of the map between a Lie algebra and its induced Lie group up to questions of covering, the physical definitions of SU(2) and SU(1,1) are often given in terms of the simpler, discrete list of elements generating the corresponding algebra (i.e., acting as a basis for the vector space indicated in Def.~\ref{def:lie_algebra}). 

\begin{definition}[SU(2)] \label{def:su2}
    The compact Lie group SU(2) can be defined by its generating algebra $\mathfrak{su}(2)$. This algebra is often given the following presentation:
        \begin{align}
            [J_x, J_y] &= iJ_z,\nonumber\\
            [J_y, J_z] &= iJ_x,\nonumber\\
            [J_z, J_x] &= iJ_y,\label{eq:su2_commutation}
        \end{align}
    where we conflate $J_x, J_y, J_z$ with $X, Y, Z$, the Pauli matrices. It is not difficult to identify the generated Lie group with the sphere (and indeed the double covering of SO(3) by SU(2) is often used in quantum computation).
\end{definition}

\begin{definition}[SU(1,1)] \label{def:su11}
    The non-compact Lie group SU(1,1) can also be defined by its generating algebra $\mathfrak{su}(1, 1)$. In a physical setting this algebra is usually given the following presentation.
        \begin{align}
            [K_x, K_y] &= -iK_z\nonumber\\
            [K_y, K_z] &= iK_x\nonumber\\
            [K_z, K_x] &= iK_y.\label{eq:su11_commutation}
        \end{align}
    As a side note, often defined are the \emph{raising} and \emph{lowering} operators $K_{\pm} = (K_x \pm i K_y)$, from which, by standard techniques in quantum mechanics, a useful complete set of basis vectors which are common eigenstates of $K_z$ and $K^2 = K_x^2 - K_y^2 - K_z^2$ are defined. Equivalently one can consider the two-by-two matrices $M$ such that
        \begin{equation}
            M^\dagger P M = P,
        \end{equation}
    where $P$ is the matrix $\text{diag}(1,-1)$, meaning that such matrices $M$ preserve $P$ up to the given product. It is not so difficult to find a parameterization of such matrices, namely
        \begin{equation}
            M(\beta, \phi, \psi) = 
            \begin{pmatrix}
                e^{i\psi}\cosh{\beta} & e^{i\phi}\sinh{\beta}\\
                e^{-i\phi}\sinh{\beta} & e^{-i\psi}\cosh{\beta}
            \end{pmatrix},
        \end{equation}
    where for our purposes, ignoring the overall phase (equivalently setting $\psi = 0$) is sufficient for our purposes. Additionally, in all realistic settings the maximum allowed $\beta$ is some finite constant, and we will often refer to this implicit bound in later results. Note that while the above is commonly given as a definition for SU(1,1), it is instead a representation of SU(1,1), with simple expression in terms of the exponential map applied to the Pauli-like generators of the algebra $\mathfrak{su}(1,1)$.
\end{definition}

The astute reader notices that SU(2) (Def.~\ref{def:su2}) and SU(1,1) (Def.~\ref{def:su11}) are almost identical in their definition, up to signs in the defining commutation relations. And indeed, the complexification of the algebras for both SU(2) and SU(1,1) are the same SL(2, $\mathbb{C}$). From the casual definition of Lie algebras as the tangent space of the idenitity of a corresponding Lie group, we see that the magnitude of the curvature of these manifolds is constant but different in sign between these two Lie groups, suggesting the hyperbolic character of SU(1,1). Both represent three-dimensional vector spaces and, as per the definition of Lie algebras, have generators which commute or anticommute among themselves.

With these basic definitions out of the way, it is possible to concretely define the types of quantum evolutions (essentially products of SU(1,1) transformations) we would like to physically realize to most closely emulate the structure of the QSP circuit ansatz. We will discuss the properties of this alternating ansatz, and then some examples of its physical realization. Along the way we will try to clarify where seemingly analagous mathematical properties between standard QSP and our modified ansatz may represent dramatically different physical properties.

\subsection{Problem statement}

Taking inspiration from the QSP constructions given in Appendix~\ref{appendix:basics_qsp_qsvt}, we can build up products of the analogous object to the phased iterate, namely the phased boost (Def.~\ref{def:phased_boost}), and compare the structures of the resulting evolutions. It is important to note, as will be remedied later, that necessarily this product represents something physically quite different than its analogue in standard QSP, foremost by the fact that the matrix considered is not an irreducible unitary representation of the evolution being studied (the existence of one for SU(1,1) being impossible under the constraint that it be finite-dimensional, as discussed in Appendix~\ref{appendix:basics_rep_theory}).

\begin{definition}[Phased boost] \label{def:phased_boost}
    The basic element of SU(1,1) used in our protocol will be the phased boost, which has the explicit form
        \begin{equation}
            V_{\phi}(\beta) \equiv 
            \begin{pmatrix}
                \cosh{\beta} & e^{i\phi}\sinh{\beta}\\
                e^{-i\phi}\sinh{\beta} & \cosh{\beta}
            \end{pmatrix},
        \end{equation}
    where $\beta \in [0, \infty)$ (with some implicit cutoff to be discussed) is called the boost parameter, and $\phi$ is some rotation angle. In terms of representations for the commonly cited generators of SU(1,1) this could be expressed by the simple conjugation
        \begin{equation}
            V_{\phi}(\beta) = e^{i\phi K_z}e^{i \beta K_x} e^{-i\phi K_z}.
        \end{equation}
\end{definition}

\begin{definition}[SU(1,1)-based QSP] \label{def:qsp_su11}
    The direct analogue of the QSP circuit ansatz \cite{lyc_16_equiangular_gates, lc_17_simultation, lc_19_qubitization, gslw_19} in SU(1,1) is the repeated application of \emph{phased boosts} (Def.~\ref{def:phased_boost}). This results in the following element of SU(1,1):
        \begin{equation} \label{eq:qsp_su11_form}
            S_{\Phi} = \prod_{k = 0}^{n} V_{\phi_k}(\beta),
        \end{equation}
    where $\Phi \in \mathbb{R}^{n + 1}$ as per usual, and $\beta \in [0, \infty)$ is generally taken to be the unknown scalar \emph{signal} being processed. We will also refer to this signal, in analogy to standard QSP, in the transformed picture $x = \cosh{\beta}$, where  $x \in [1, \infty)$, as in Eq.~\ref{eq:hyperbolic_form}. As in standard QSP, an evolution of this form represents a map from a list of real numbers to a function from a (artificially constrained) compact subset of possible $\beta$ to elements of SU(1,1).
        \begin{equation} \label{eq:hqsp_type_definition}
            \text{SU(1,1)-QSP} : \mathbb{R}^n \rightarrow (F: [-\beta, \beta] \rightarrow SU(1,1)).
        \end{equation}
    Here we have overloaded $\beta$ to refer also to largest permitted boost parameter in our ansatz. It will turn out that finite energy and experimental constraints will always place a reasonable limit on the allowed $\beta$.
\end{definition}

For the moment we do not discuss the physical relevance of the product of phased boosts given in Def.~\ref{def:qsp_su11}. Instead, we investigate the formal properties of this product, the map it induces to a space of functions, and the extent to which standard statements in QSP port to this new picture. It turns out that the similarities of the defining algebras (Defs.~\ref{def:su2} and \ref{def:su11}) permit plenty to be directly learned from the theory of standard QSP re this mathematical object. However, as the astute reader will remember that non-compact Lie groups do not permit finite-dimensional unitary representations, the matrices we will consider will not represent, as was the case in QSP, the evolution of quantum states, but rather (for our purposes) that of quantum operators.

\begin{theorem}[Functional form of SU(1,1) QSP] \label{thm:hyperbolic_substitution}
    The achievable functional form of a SU(1,1) QSP protocol with constituting phases $\Phi$ is similar to that of a standard QSP protocol, namely the product in Eq.~\ref{eq:qsp_su11_form} permits the non-unitary finite dimensional representation
        \begin{equation} \label{eq:hyperbolic_form}
            S_{\Phi} = 
            \begin{pmatrix}
                P & Q\sqrt{x^2 - 1} \\
                Q^*\sqrt{x^2 - 1} & P^*
            \end{pmatrix},
        \end{equation}
    with $P, Q$ having definite (but opposite) parity in $x$, degree bounded above by the length $(n + 1)$ of $\Phi \in \mathbb{R}^{n + 1}$, and satisfying $|P|^2 - (x^2 - 1)|Q|^2 = 1$, up to the replacement $x = \cosh{\beta}$, in analogy with standard QSP. Proof of this statement is by direct substitution, observing the forms of the phased iterates in standard QSP and SU(1,1) QSP, reproduced respectively below.
        \begin{equation} \label{eq:phased_iterates}
            W_{\phi}(\theta) \equiv 
            \begin{pmatrix}
                \cos{\theta} & e^{i\phi}\sin{\theta}\\
                -e^{-i\phi}\sin{\theta} & \cos{\theta}
            \end{pmatrix},
            \quad
            V_{\phi}(\beta) \equiv 
            \begin{pmatrix}
                \cosh{\beta} & e^{i\phi}\sinh{\beta}\\
                e^{-i\phi}\sinh{\beta} & \cosh{\beta}
            \end{pmatrix}.
        \end{equation}
    Note specifically that for standard QSP, with $x = \cosh{\theta}$, that the analytic continuation of $\theta \mapsto i\theta$ takes $\cos{\theta}$ to $\cosh{\theta}$ and $\sin{\theta}$ to $i\sinh{\theta}$, and consequently $\sqrt{1 - x^2}$ to $\sqrt{x^2 - 1}$. By modifying the angles defining the phased iterate in QSP, namely
        \begin{equation}
             W_{[\phi - \pi/2]}(\theta) \equiv 
            \begin{pmatrix}
                \cos{\theta} & -e^{i\phi}i\sin{\theta}\\
                -e^{-i\phi}i\sin{\theta} & \cos{\theta}
            \end{pmatrix},
        \end{equation}
    we see that the analytic continuation $\theta \mapsto i\theta$ precisely recovers the form of the phased boost used in SU(1,1) QSP. In this sense the polynomials in $x = \cosh{\beta}$, \emph{as defined by their coefficients}, are in bijection with those of standard QSP. Note that this does not immediately imply that these polynomials are dense in a useful space of functions; this needs to be shown, and relies on a number of basic results in the theory of functional approximation.
\end{theorem}

\begin{remark}[On analytically continuing the QSP ansatz] \label{remark:analytic_continuation}
    As discussed in Theorem~\ref{thm:hyperbolic_substitution}, there are a few implicit maps between SU(2) and SU(1,1) based QSP. First, one can interpret that the signal oracle has been transformed to rotate by an imaginary angle:
        \begin{equation}
            \theta \mapsto i\theta = \beta.
        \end{equation}
    In this sense, when viewing QSP unitaries as embedding Laurent polynomials (as in \cite{haah_2019, rc_22}, the function is here being evaluated \emph{off the unit circle}. Equivalently, taking the natural assignment
        \begin{equation}
            x \equiv \cos{\theta} \mapsto x \equiv \cosh{\beta},
        \end{equation}
    the transformation can instead be seen as evaluating QSP-induced polynomial transforms outside the range $x \in [-1,1]$. We will often work in the latter picture to follow previous convention, but both indicate that we now seek to control the behavior of these polynomials outside of regions previously considered (and thus under novel constraints).
\end{remark}

It is quick work, once the bijection between QSP protocols and SU(1,1) QSP protocols has been established, to prove an analogous theorem to that in \cite{gslw_19}, namely that a \emph{partially specified} finite dimensional representation for a SU(1,1) QSP protocol can be \emph{completed}, that is, its missing elements filled in and its corresponding phases read off following the standard techniques of QSP \cite{chao_machine_prec_20, haah_2019, dong_efficient_phases_21}.

\begin{theorem}[Matrix completion in SU(1,1) QSP] \label{thm:matrix_completion_su11}
    Take $\beta \in [-\gamma, \gamma]$ for some $\gamma \in \mathbb{R}$ and consider polynomials $P, Q \in \mathbb{R}[\cosh{\beta}]$ where $x \equiv \cosh{\beta}$ such that the following conditions hold
        \begin{enumerate}
            \item $P$ has degree $n$ and $Q$ has degree $n - 1$
            \item $P$ has parity $n \pmod{2}$ and $Q$ has parity $(n-1) \pmod{2}$.
            \item $P^2 - (x^2 - 1)Q^2 \geq 1$ for $x \in [1, \infty)$.
        \end{enumerate}
    Then there exists $\Phi \in \mathbb{R}^{n+1}$ such that the SU(1,1) QSP protocol with phases $\Phi$ (Def.~\ref{def:qsp_su11}) has the form
        \begin{equation}
            S_{\Phi} = 
            \begin{pmatrix}
                \tilde{P} & \tilde{Q}\sqrt{x^2 - 1} \\
                \tilde{Q}^*\sqrt{x^2 - 1} & \tilde{P}^*
            \end{pmatrix},
        \end{equation}
    where $\Re[\tilde{P}] = P$ and $\Re[\tilde{Q}] = Q$. The phases $\Phi$ are known to be efficiently computable in time $\text{poly}(n)$ by a classical algorithm \cite{dong_efficient_phases_21, haah_2019}. Both of these results follow from the bijection with QSP up to analytic continuation in the variable $\theta$. Making the reverse substitution $\beta \mapsto -i\beta$ and $\phi \mapsto \phi + \pi/2$ allows the required phases to be recovered and repurposed for the SU(1,1) QSP protocol.
\end{theorem}

The bijection between \emph{circuit descriptions} and \emph{polynomial coefficients} for standard QSP versus SU(1,1) QSP is promising when considering their similarities, but note that this does not tell the entire story. The space of functions on the right-hand-side of the type definition for SU(1,1) QSP protocols given in Eq.~\ref{eq:hqsp_type_definition} is currently not well-defined. Indeed, the map between both circuit descriptions and unitaries, as well as polynomial coefficients and functions of the underlying signal parameter $\beta$, is not injective, let alone obviously preserving of functional analytic properties like density in a space of functions. We leave this discussion for Sec.~\ref{sec:analyzing_qsp_ansatz}, devoted to functional approximation theory. Before this, however, we discuss one physical instantiation of SU(1,1) interactions in quantum information, which will serve as a model for intuition going forward.

\subsection{A simple physical implementation}

\begin{figure}
    \centering
    \includegraphics[width=0.6\textwidth]{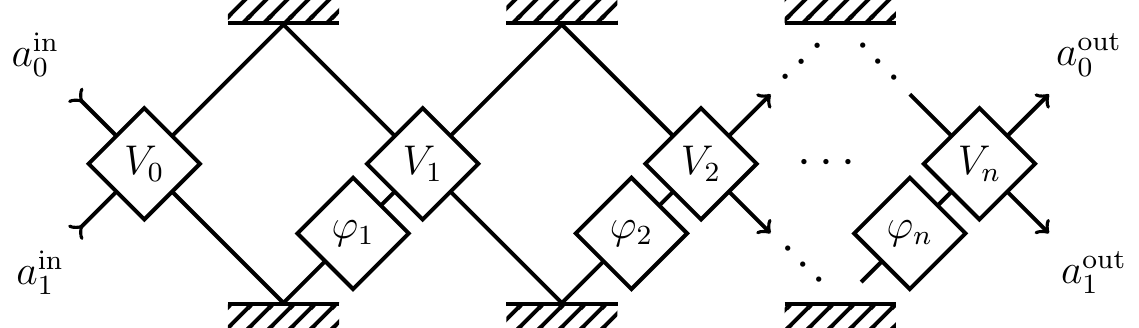}
    \caption{One instantiation of a staged SU(1,1) interferometer, following the model considered in \cite{ou_li_su11_review_20}, using notation from \cite{chuang_simple_computer_95}. Here two modes are fed into a series of four-wave-mixing parametric amplifiers, where a phase shift $\phi_k$ is applied to one arm of the interferometer between each amplification $V_k$ (as per $V_\phi$ for $\phi = 0$ in Eq.~\ref{eq:phased_iterates}). Dashed boxes are mirrors, while lines represent the path of light, moving from left to right. The output modes depend non-linearly on the underlying parameters of each parametric amplification; these unknown amplifications, as in standard QSP, are assumed to be \emph{consistent}: the same each time applied.}
    \label{fig:su11_interferometer}
\end{figure}

In this section we ground SU(1,1) in the physical context of interferometry. Again we aim to present this minimally, pointing the interested reader toward recent in-depth experimental work in this topic, extensive examination of which is beyond the scope of this paper. We also note that even within the two major interferometric regimes we discuss (i.e., beam-splitter and parametric amplifier based apparatuses) there are multiple incomparable devices with differing performance and underlying physical mechanisms. Ultimately, the primary goal of this section, as shown in Fig.~\ref{fig:su11_interferometer}, is to realize the dynamics given in Def.~\ref{def:qsp_su11} by a reasonable physical device. For us this means specifying the proper (in this case optical) element achieving the SU(1,1) phased iterate of Eq.~\ref{eq:phased_iterates}.

\begin{definition}[SU(1,1) interferometer] \label{def:su11_pa_hamiltonian}
    For the rest of this work we refer to SU(1,1) interferometers as those whose primary optical element is a parametric amplifier. The defining interaction term for this Hamiltonian is the following
        \begin{equation}
            H = i\hbar \xi a_1^\dagger a_2^\dagger + \text{h.c.},
        \end{equation}
    where we will usually refer to the composite variable $\beta$ which will be proportional to $\xi$ and thus the nonlinear coefficient and pump field amplitudes that are producing the intended amplification. The term $\beta$ can be easily used to define the evolutions of modes under this Hamiltonian, as discussed below.
\end{definition}

SU(1,1) interferometry can be summarized as an ability to perform the following mode transformations, induced by the interaction Hamiltonian in Def.~\ref{def:su11_pa_hamiltonian}. We mainly follow the notation of the work introducing SU(1,1) interferometry \cite{yurke_su2_su11_86}, which in our simple setting is more than sufficient.
    \begin{align}
        a_1 &\mapsto (a_1)\,\cosh{\beta} + (a_2^{\dagger})\,e^{i\phi}\sinh{\beta},\nonumber\\
        a_2 &\mapsto (a_2)\,e^{-i\phi}\sinh{\beta} + (a_1^\dagger)\,\cosh{\beta}.\label{eq:su11_mode_map}
    \end{align}
In other words, the phased boost discussed in Def.~\ref{def:phased_boost} describes the manipulation of the operators $a_1, a_2$ and their complex conjugates under the action of a series of parametric amplifications and phase shifts. Conjugating a squeezing operation by phase shifts is precisely the statement of Def.~\ref{def:phased_boost}.

\begin{definition}[Staging SU(1,1) interferometers]
    The mode transformations enacted by both beam-splitters and parametric-amplifier based interferometers can be applied sequentially to achieve more complicated transformations of the mode. Such a sequence will be referred to as a staged SU(1,1) interferometer, and is depicted in Fig.~\ref{fig:su11_interferometer}. In its simplest form, however, as discussed in \cite{ou_li_su11_review_20}, one can consider the low-gain regime, where $\beta$ is small, and therefore the applied unitary (now in the Schrödinger picture) has the simple approximation
        \begin{equation}
            U \approx I + (e^{i\phi}\sinh{\beta}\,a_1^\dagger a_2^\dagger + \text{h.c.}) + \mathcal{O}(\sinh^2{\beta}),
        \end{equation}
    in which case repeated applications of $U$ interspersed with some phase shift unitary $\Theta$ inducing a phase difference of $\theta$ between the two modes will result in the following transformation (again in Schrödinger picture) after application $N$ times:
        \begin{equation}
            |00\rangle \mapsto |00\rangle + e^{i\phi}\sinh{\beta} \left[\sum_{k = 1}^{N} e^{-i(k - 1)\theta}\right] |11\rangle + \mathcal{O}(\sinh^2{\beta}).
        \end{equation}
    It is worth noting that in this limit, we indeed simply have the effective action of one SU(1,1) interferometric operation with modified $\beta$ term corresponding to the following substitution:
        \begin{equation}
            e^{i\phi}\sinh{\beta} \mapsto e^{i(\phi - [N - 1]\theta/2)} \sinh{\beta} \frac{\sin{[N\theta/2]}}{\sin{[\theta/2]}}.
        \end{equation}
    We note that this does not make use of the non-commutative aspect of the Hamiltonian's various terms, and in fact that this approximation restricts us to a qubit-like subspace spanned by the vacuum state and $|1,1\rangle$. It is the goal of this work to explicitly violate some of the assumptions of this approximation, which necessarily takes us out of a nice, finite dimensional unitary representation for system dynamics.
\end{definition}

While the evolution of the modes in SU(1,1) interferometry matches the mathematical formalism of the previous section, the primary utility of QSP as originally introduced lies in the polynomial transformation of an unknown signal. Taking the parameter $\beta$ or $\cosh{\beta}$ as the unknown, it is worthwhile to consider settings in which this value can be both (1) reasonably said to be unknown, or otherwise (2) varying with respect to some other degree of freedom such that the overall transformation, non-linearly dependent on this unknown, performs a useful task. We discuss one possible concrete method of coupling SU(1,1) operations to external quantum systems, such that a protocol as in Fig.~\ref{fig:su11_interferometer} could be implemented. A major open question along this line of work is whether one can construct further natural couplings between (possibly unknown) systems and SU(1,1) interferometric operations.

\begin{definition}[Controlled-squeezing operations]
    A controlled-squeezing operation considers the case where the mode-transformation discussed previously is coherently controlled by the state of another quantum system. For qubit-coupled oscillators, a simple interaction could take exactly this form, e.g.,
        \begin{equation}
            C(\beta_0, \beta_1) 
            \equiv
            |0\rangle\langle 0|\otimes W_{\phi}(\beta_0) + |1\rangle\langle1|\otimes W_{\phi}(\beta_1).
        \end{equation}
    Consequently the composite map between modes defined by products of the phased boost used in SU(1,1) QSP, i.e.,
        \begin{align}
            a_1 &\mapsto (a_1)\,P(\cosh{\beta}) + (a_2^\dagger)\,[\sinh{\beta}]\,Q(\cosh{\beta}) ,\\
            a_2 &\mapsto (a_2)\,[\sinh{\beta}]\,Q^*(\cosh{\beta}) + (a_1^\dagger)\,P^*(\cosh{\beta}),
        \end{align}
    can be applied in superposition according to the quantum state of some auxiliary, perhaps qubit-based system. This sort of coupling could be advantageous in systems where the physical instantiation of the auxiliary qubit is one more amenable to measurement, while the system on which the parametric amplifier acts is more resistant to noise. Such problems of coupling are ubiquitous in QSP and QSVT, where they form the basis of the theory of block-encoded linear operators.
\end{definition}

For controlled-squeezing operations, the utility of a QSP-like ansatz is clear: one can perform these QSP manipulations in superposition according to the state of the coupled qubit. Indeed, given the indefinite direction of controlled operations in quantum computing, such circuits can also be used to preferentially prepare the coupled qubit into a desired state based on the magnitude or direction of the squeezing operation. For the moment though, we set the question of optimal physical instantiation of this method alone, and focus instead on the expressivity of the SU(1,1) QSP ansatz in comparison to its standard cousin.

\section{The expressivity of SU(1,1) QSP} \label{sec:analyzing_qsp_ansatz}

As mentioned the main difference between the theory of standard QSP and SU(1,1) analogue is the replacement of standard trigonometric polynomials with their hyperbolic equivalents (Remark~\ref{remark:analytic_continuation}). In this setting, many of the basic results of functional approximation theory employed in standard QSP are no longer immediately applicable. This section discusses methods by which the functional expressivity of QSP-like algorithms are proven, and interprets relevant theorems for SU(1,1) QSP. As in standard QSP, many of the theorems we care about are, in their most generic form, already fundamentally known in functional analysis and approximation theory: the business of QSP is often massaging our problem statement and use case into a form that matches these established theorems' assumptions. We also discuss the numerical efficiency of classical subroutines used to specify QSP protocols (i.e., those algorithms to compute QSP phases to a specified precision given a desired embedded polynomial). Unless otherwise noted, standard definitions and theorems in topology and functional analysis are taken from common textbooks \cite{bourbaki_66, rudin_func_91}.

\subsection{The Stone-Weierstrass theorem}

We briefly cover results related to functional approximation, the most famous of which concern the approximation of functions by polynomials. Toward discussing approximation with the natural functions embeddable in SU(1,1) QSP, we provide a series of definitions and standard results. It should be noted that this section mainly discusses how one determines whether a set of functions generates a sub-algebra of the set of continuous functions that is \emph{dense} in the set of continuous functions. This is often removed from discussing the efficiency of such approximation or the ease of its computation.
    
\begin{definition}[Hausdorff space] \label{def:hausdorff_space}
    A Hausdorff space $X$ is a topological space such that for any two distinct elements, $x_1, x_2$, there exist two open sets, $U_1, U_2$, such that $x_1 \in U_1, x_2 \in U_2$ and $U_1 \cap U_2 = \emptyset$.
\end{definition}

\begin{definition}[Uniform metric and convergence] \label{def:uniform_metric}
    A metric space $C(X, \mathbb{R})$ is said to have the \emph{uniform metric} if the distance between two functions $f, g$ is computed according to
        \begin{equation}
            d(f, g) \equiv \sup_{x \in X} \lvert f(x) - g(x) \rvert.
        \end{equation}
    Given a sequence of functions $f_0, f_1, \cdots$, this sequence is said to \emph{uniformly converge} to some $g$ if the sequence of real numbers $d(f_n, g)$ converges to zero. Sometimes a space of functions is said to have \emph{the topology of uniform convergence} if its underlying metric space is equipped with the uniform metric.
\end{definition}

\begin{theorem}[Stone-Weierstrass theorem (general form)] \label{thm:stone_weierstrass}
    Suppose $X$ is a compact Hausdorff space and $A$ is a subalgebra of $C(X, \mathbb{R})$ (real-valued continuous functions on $X$ with the topology induced by $\lVert \cdot \rVert_\infty$) which contains a non-zero constant function. Then $A$ is dense in $C(X, \mathbb{R})$ if and only if it separates points, i.e., that for every $x \neq y \in X$ there exists some function $p \in A$ such that $p(x) \neq p(y)$.
\end{theorem}

\begin{lemma}[Hyperbolic trigonometric functions as bases] \label{lemma:complete_basis}
    The hyperbolic trigonometric functions $S = \{\cosh{(nx)}\}$ for $n \in \mathbb{N}$ form a complete basis for square integrable functions with compact support on some interval $x \in [-c, c]$ for $c \in \mathbb{R}$. Proof is by bootstrapping via the Stone-Weierstrass theorem. It can easily be seen that the set of real exponentials $\{e^{n\theta}\}, n \in \mathbb{Z}$ defined on a finite interval $[-\beta, \beta], \beta > 0$ both separates points (in fact they do not overlap beyond at $\beta = 0$) and contains a non-zero constant function, namely $1$. Note that this property is not modified if one considers hyperbolic trigonometric functions or real exponential functions, again indexed by the integers.
\end{lemma}

\begin{definition}[$L^2$ functions, $L^p$ spaces, and integrability] \label{def:l2_space}
    A square integrable function, equivalently an $L^2$-function is a real or complex valued function for which the integral of the square of its absolute value is finite. On the real line this is the statement, for $L^2$-function $f$, that
        \begin{equation}
            \int_{-\infty}^{\infty} |f(x)|^2 \,dx < \infty.
        \end{equation}
    It is also equivalent to say that the square of the absolute value of the function is Lesbegue integrable. The vector space of square integrable functions is called $L^2$, for which the extension to arbitrary positive integers $p$ define the $L^p$ spaces. The space $L^2$ is unique among the $L^p$ spaces for being compatible with an inner product among functions, and we consider it exclusively.
\end{definition}

\begin{lemma}[Density in square-integrable functions] \label{lemma:density_in_l2}
    The set $C([0, 1], \mathbb{R})$ is dense in $L^2[0, 1]$ (the space of square, Lebesgue-integrable functions). Proof is standard in functional analysis. This permits us to use Stone-Weierstrass results for approximating square integrable functions, which are the common goal in Fourier analysis. By the applicability of Stone-Weierstrass to the hyperbolic trigonometric functions we are also able to approximate square integrable functions.
\end{lemma}

\begin{remark}[On the efficient approximation of a desired function with an arbitrary dense sub-algebra of continuous functions] \label{remark:parsevals_thm}
    We know from suitably modified versions of Parseval's theorem \cite{rudin_func_91} the generalized Fourier coefficients for the approximation to a given square-integrable function have the sum of their squared-magnitudes equal to the result of the integral of the square of the magnitude of the function itself on the relevant interval (this is one statement of the unitarity of the Fourier transform). In other words, given two complex-valued square-integrable functions $f, g$ over the reals
        \begin{equation}
            \sum_{n = -\infty}^{n = \infty} f_n g_n = 
            \frac{1}{2\pi}\int_{-\infty}^{\infty} f(x)g(x)\,dx,
        \end{equation}
    where $f_n, g_n$ are the $n$-th Fourier coefficients of $f$ and $g$ respectively. Taking $f = g$ and assuming $f$ to be square integrable, this is a statement that the square integrability of a function is equivalent to a finite sum of the squared-magnitudes of its Fourier coefficients. It is known that this relation holds even for generalized Fourier coefficients induced by any valid choice of a complete orthogonal system of univariate functions, and so will notably also hold in our case.
\end{remark}

\begin{remark}[On Gram-Schmidt orthgonalization] \label{remark:gram_schmidt}
    In standard QSP the basically achieved functions, e.g., $\langle 0 |U_{\Phi} |0\rangle$, for trivial QSP phase lists $\{0, 0, \cdots, 0\}$ are the Chebyshev polynomials, which are naturally orthogonal on $[-1,1]$ with respect to the simple functional inner product
        \begin{equation}
            \langle P, Q\rangle = \int_{-1}^{-1} P(x)Q(x)\,\frac{dx}{\sqrt{1 - x^2}}.
        \end{equation}
    In SU(1,1) QSP, the same polynomials are now evaluated over a different region, $[1, \cosh{\beta}]$ for some finite positive $\beta$, and the orthogonality enjoyed by the Chebyshev polynomials is lost. Nevertheless, the density of these polynomials in the relevant functional space discussed above is preserved, and thus successive Gram-Schmidt orthonormalization of these functions is possible, albeit possibly numerically ill-conditioned with increasing degree. We can choose to implicitly work in one of these manufactured bases when discussing Fourier analysis in the next section.
\end{remark}

\subsection{On non-harmonic Fourier analysis} \label{sec:non_harmonic_analysis}

In the previous section we focused on, for the SU(1,1)-variant ansatz of QSP, the density of generated transforms in a reasonable space of functions. While such density results are useful and necessary for the application of QSP, they are not sufficient. More specifically, one of the major benefits of standard QSP is that the embedded functional transform can be made to quickly converge to a desired functional transform, implying that the realizing circuit is relatively short. I.e., while we have a good description of the types of polynomials that are permissible in these transforms from Theorem~\ref{thm:matrix_completion_su11}, the speed with which such polynomials converge to desired continuous functions with the same properties is not obvious. In other words, we also seek to relate properties of the desired function to the minimum required length of interleaving ansätze whose embedded polynomial transforms achieving such a function (up to uniform approximation).

As discussed, in standard QSP the achieved functional transforms are trigonometric polynomials in $\cos{\theta}$, which have clean Fourier series. The nice property we care about in relation to these implicit complex exponentials $\{e^{i n \theta}\}, n \in \mathbb{Z}$ is that they are \emph{closed} over $[-\pi, \pi]$, namely that for Lebesgue-integrable functions $f(\theta)$, the equation
    \begin{equation}
        f_n = \int_{-\pi}^{\pi} f(\theta)\,e^{in\theta}\,d\theta = 0,
    \end{equation}
holding for all $n \in \mathbb{Z}$ implies that $f(\theta) = 0$ identically. This is one of the basic observations of Fourier analysis, relating closely to the unitarity of the Fourier transform. Integrals similar to those above yield a set of complex numbers $f_n$ (the Fourier coefficients), which can be used to well-approximate a wide class of desired functions. We would like to recover similar properties for SU(1,1) QSP, as well as connect this discussion (on closed sets of functions) to the previous section on density in useful classes of functions.

In literature the results we care about are referred to as closure and gap theorems. These theorems seek to assert similar statements to those applied in standard Fourier analysis, save one considers a possibly non-orthogonal basis of functions $\{e^{i \lambda_n \theta}\}, n \in \mathbb{Z}$ with possibly complex $\lambda_n$. Gap theorems specifically consider families of functions for which coefficients corresponding to some set $\{\lambda_n\}$ are zero, in which case one may be able to assert that, only on some subinterval, all functions which have vanishing coefficients can again only be the function which is identically zero on that interval. In general however we are mainly concerned with closure theorems, which are strong enough for the purposes of this work.

We now show that completeness and closure (properties of sets of functions on intervals) are for us effectively interchangeable. We then show that a small modification to a known complete/closed set of functions (complex exponentials) yields a function set which (1) aligns with that of SU(1,1) QSP and (2) maintains desired closure completeness properties. We will exclusively work with $L^2$ (square-integrable) functions, unless otherwise noted. Toward this result, we cite a number of constitutive definitions and theorems.

\begin{definition}[Closed {$L^p[-a, a]$} function set] \label{def:closure}
    A set of function $\{f_n\}, n \in \mathbb{N}$ is $L^p$ closed on an interval $[-a, a], a > 0$ if for every function $g \in L^p[-a, a]$, $g$ can be approximated according to the $L^p$ norm by linear combinations of $f_n$ with possibly complex coefficients.
\end{definition}

\begin{remark}[On closed versus complete function sets] \label{remark:closure_vs_completeness}
    A set of functions is said to be incomplete in $L^p$ if there exists a non-trivial function in $L^p$ which is orthogonal to all function in that set. Closure, however, makes a statement about the \emph{approximation} of functions in $L^p$ (with respect to some interval). For general $p$ and over general measure spaces, the equivalence of closure and completeness is neither obvious nor necessarily true \cite{young_non_harmonic_01}.
\end{remark}

\begin{definition}[Measure space] \label{def:measure_space}
    A measure space is a tuple $(X, \mathcal{A}, \mu)$ of a set $X$, a $\sigma$-algebra $\mathcal{A}$ for $X$ and a measure $\mu$ on $(X, \mathcal{A}$ (the latter common called a measurable set). Here $\mathcal{A}$ is used to assign measurability to $X$, while $\mu$ is used for computing the size of various subsets of $X$. We also say a measure $\mu$ is $\sigma$-finite if it is a countable union of measurable sets with finite measure, i.e., $\mu(X_k) < \infty$, where $\cup X_k = X$ (a common example is the Lebesgue measure over the reals, which is not finite, but which is $\sigma$-finite).
\end{definition}

\begin{theorem}[Completeness and closure equivalence for $L^2$ functions (Example 1 in \cite{young_non_harmonic_01})] \label{thm:completeness_closure_equivalence}
    If $(X, \mathcal{A}, \mu)$ is a $\sigma$-finite measure space and $1 \leq p < \infty$, then the Riesz representation theorem shows that the dual of $L^p(\mu)$ can be identified with $L^q(\mu)$, where $1/p + 1/q = 1$. From this it follows that a sequence of functions $\{f_n\}, n \in \mathbb{N}$ in $L^p(\mu)$ is closed over $L^p[X]$ if it is complete over $L^p[X]$. For $p = 2$, the space of $L^2$ functions over $X$ equipped with the measure $\mu$ is self-dual, and closure and completeness coincide.
\end{theorem}

We now cite and apply a few results from foundational work in functional analysis on closure theorems, geared toward our specific setting. The main cited sources include a standard textbook on non-harmonic analysis \cite{young_non_harmonic_01}, as well as some of the monographs it cites, which provide full proofs of the provided statements \cite{levinson_gap_40, levinson_36}. Where relevant we also cite more recent work \cite{redheffer_completeness_77, redheffer_completeness_83}, which supply streamlined proofs.

\begin{theorem}[Theorem 4 in \cite{levinson_gap_40}, cited as such in \cite{young_non_harmonic_01} (Theorem 4) and \cite{redheffer_completeness_77, redheffer_completeness_83} (Theorem 9) in generalized forms] \label{thm:levinson_completeness}
    If $0 < p < \infty$ and $\{\lambda_n\}$ is a sequence of real or complex numbers for which
        \begin{equation}
            |\lambda_n | \leq |n| + \frac{1}{2p}, \quad n \in \mathbb{Z},
        \end{equation}
    then the system $\{e^{i\lambda_n \theta}\}$ is complete in $L^{p}[-\pi, \pi]$, and the term $1/2p$ cannot be improved in general. Consequently by Theorem~\ref{thm:completeness_closure_equivalence}, for $p = 2$ this functional set is also closed in $L^{2}[-\pi, \pi]$.
    
    Proof of this statement given in terms of properties of a counting function. Concretely, this is a statement that shows that a set $\{e^{i\lambda_n\theta}\}, n \in \mathbb{N}$ is complete for $L^p$ on an interval of length $2\pi D$ if
        \begin{equation} \label{eq:root_counting}
            \limsup_{r \rightarrow \infty} \left(\int_{1}^{r} \frac{\Lambda(\theta) - 2 D \theta}{\theta}\,d\theta + \frac{\log{q}}{r}\right) > -\infty.
        \end{equation}
    Here $\Lambda(\theta)$ is the number of points in $\{\lambda_n\}$ (complex) inside the disk of radius $\theta$, referred to as an unsigned counting function.
\end{theorem}

The proof method cited briefly in Theorem~\ref{thm:levinson_completeness}, and its accompanying imposed condition in Eq.~\ref{eq:root_counting}, can appear quite abstract, and so we take a moment to discuss its heuristic justification. In fact, as discussed in \cite{young_non_harmonic_01}, proving the completeness of sets of functions by investigating the properties of roots of special functions is an exceedingly common technique (if not often the only commonly employed technique). The expression in Eq.~\ref{eq:root_counting} permits proof in the following way. Assume toward contradiction of the completeness of the $f_n$ that all $F(\lambda_n)$ (the $n$-th generalized Fourier coefficients of some $f$) are zero. Then the following inequality holds
    \begin{equation}
        |F(z)| \leq \int_{-a + \delta}^{a - \delta}e^{-yt}|f(t)|\,dt + \int_{\text{out}}e^{-yt}|f(t)|\,dt,
    \end{equation}
where $z = x + iy$ and $\delta > 0$ is some small positive number, and `out' refers to the portions of the interval beyond $a - \delta$ and before $-a + \delta$. Hölder's inequality states that if $\lVert f \rVert$ is small then
    \begin{equation}
        |F(z)| \leq e^{a|y|}|y|^{1/q}(e^{-\delta |y|} + \eta),
    \end{equation}
where $\eta$ goes to zero as $\delta$ does. This permits us to write out the following integral inequality, following the simplified calculations in Theorem 8 of \cite{redheffer_completeness_77}:
    \begin{align}
        \int_{-\pi}^{\pi} \log\left[F(re^{i\theta})\right]\,d\theta  &\leq
        \int_{-\pi}^{\pi}ar|\sin{\theta}|\,d\theta - (1/q)\int_{-\pi}^{\pi}\log{r}\,d\theta 
        - (1/q)\int_{-\pi}^{\pi}\log{|\sin{\theta}|}\,d\theta \nonumber\\
        &+ \int_{-\pi/3}^{\pi/3}\log\left(e^{\delta r/2} + \eta\right)\,d\theta  + \int_{\text{out}} \log{\left(1 + \eta\theta\right)}\,d\theta.
    \end{align}
The first two integrals are simple, the third is convergent, but the fourth can be made less than any chosen negative number by first choosing $\delta$ so that $\eta$ is small, and then taking $r$ as large as necessary. This permits us to write the following inequality
    \begin{equation} \label{eq:magnitude_inequality}
        \int_{-\pi}^{\pi} \log\left[F(re^{i\theta})\right]\,d\theta \leq 2Dr - ([\log{r}]/q) - \varphi(r),
    \end{equation}
where $\varphi(r)$ goes to infinity as $r$ goes to infinity. The final step is an application of Jensen's formula in Lemma~\ref{lemma:jensens_formula}, which relates the integral we're considering to a root counting function, namely
    \begin{equation} \label{eq:jensens_inequality}
        \int_{r}^{\infty} \frac{\Lambda(t)}{t}\,dt \leq \frac{1}{2\pi}\int_{-\pi}^{\pi}\log{|F(r e^{i\theta})|}\,d\theta.
    \end{equation}
We can then see that substituting the inequality in Eq.~\ref{eq:jensens_inequality} into Eq.~\ref{eq:magnitude_inequality} shows that if the integral in Eq.~\ref{eq:root_counting} is greater than $-\infty$, that the integral of the log of $F(z)$ will be taken to negative infinity by the behavior of $\varphi(r)$, in which case the function $f$ itself will be forced to zero identically, in contradiction of our assumption of non-completeness of the $f_n$. Consequently the satisfaction of Eq.~\ref{eq:root_counting} can be used to show the completeness of the functional set. Below we cite Jensen's inequality (or formula) in more detail.

\begin{lemma}[Jensen's formula \cite{young_non_harmonic_01}] \label{lemma:jensens_formula}
    If $f(z)$ is analytic in $|z| < R$, then we denote by $\Lambda(r)$ for $0 \leq r < R$ the number of zeros $z_1, z_2, z_3, \cdots$ of $f(z)$ for which $|z_k| \leq r$. Provided that $f(0) \neq 0$, simple results in complex analysis can be used to show
        \begin{equation}
            \sum_{|z_k| \leq r} \log{\frac{r}{|z_k|}} = 
            \int_{0}^{r} \frac{\Lambda(\theta)}{\theta}\,d\theta,
        \end{equation}
    in which case Jensen's formula can be modified from its original statement to the slightly more useful
        \begin{equation}
            \frac{1}{2\pi}\int_{0}^{2\pi} \log{|f(r e^{i\phi})|}\,d\phi =
            \log{|f(0)|} + \int_{0}^{r}\frac{\Lambda(\theta)}{\theta}\,d\theta.
        \end{equation}
    This provides a concrete relation between the growth of an entire function and the density of its zeros and can be used, as shown above, in the discussion of the completeness of non-trigonometric functional bases.
\end{lemma}

The result given in Theorem~\ref{thm:levinson_completeness} discusses completeness, and by merit of Theorem~\ref{thm:completeness_closure_equivalence}, also closure of our desired ansatz in the relevant space of functions. The original question of this section, however, concerns also the efficiency of functional approximation in terms of the number of terms before truncation in order to achieve a given degree of uniform approximation. To investigate such properties, we go through yet another common object in the study of generalized Fourier analysis: Riesz bases. We provide a common definition and related theorem below.

\begin{definition}[Riesz basis; from \cite{young_non_harmonic_01}] \label{def:riesz_bases}
    A basis for a Hilbert space is a Riesz basis if it is equivalent to an orthonormal basis; that is, it is obtained from an orthonormal basis by means of a bounded invertible operator.
\end{definition}

\begin{theorem}[Properties of Riesz bases \cite{young_non_harmonic_01}] \label{thm:riesz_properties}
    Let $H$ be a separable Hilbert space; then the following are equivalent.
        \begin{enumerate}
            \item The sequence $\{f_n\}$ forms a Riesz basis for $H$.
            \item There is an equivalent inner product on $H$ for which $\{f_n\}$ becomes an orthonormal basis for $H$.
            \item The sequence $\{f_n\}$ is complete in $H$ and there exist positive constants $A, B$ such that for an arbitrary positive integer $n$ and arbitrary scalars $c_1, c_2, \cdots, c_n$ one has
                \begin{equation}
                    A \sum_{k = 1}^{n} |c_k|^2 \leq
                    \left\lVert \sum_{k = 1}^{n} c_k f_k\right\rVert \leq 
                    B \sum_{k = 1}^{n} |c_k|^2.
                \end{equation}
        \end{enumerate}
\end{theorem}

We can make a couple statements based off Definition~\ref{def:riesz_bases} and corresponding Theorem~\ref{thm:riesz_properties} of equivalent conditions. The first is that the real exponential functions $\{e^{n\theta}\}, n \in \mathbb{Z}$ \emph{do not} form a Riesz basis, despite their completeness and closure in $L^2[-\beta, \beta]$ for any finite $\beta > 0$ as shown, simply because they are not bounded on the interval $[-\beta, \beta]$ for arbitrary $n$. More intuitively, for any fixed $\beta$, the growth of the maximum of $f_n$ on the interval $[-\beta, \beta]$ is unbounded as $n$ gets large, and consequently the behavior of an embedded polynomial in $\cosh{(n\beta)}, n \in \mathbb{N}$ for large $n$ will be dominated by the leading term near $\beta$. As we have completeness and closure, we are not precluded from attempting to approximate desired functions using this ansatz, but we will have to remain careful about asymptotic statements, as there may exist sub-classes of functions for which even very long lists of Fourier coefficients approximate the desired function poorly on some sub-interval. To be fair we did not expect that these unbounded functions would constitute a Reisz basis as stated, and we provide some later indication that to assume the universal efficiency of approximation for functions outside $x \in [-1,1]$ is to assume various unphysical properties for the underlying quantum system's evolution. In what follows, however, we determine that even without this well-conditioned basis, various useful functions can nevertheless be approximated relatively quickly, with experimental utility.

\section{Worked numerical examples} \label{sec:worked_numerical_examples}

Furthering the analogy that SU(1,1)-based QSP can be viewed as standard QSP phase rotations interspersed by an oracle unitary rotating by a complex angle, the numerical side of the computation of these phases follows similar steps. In this section we provide a few examples of concretely computed phase sequences with easily interpreted actions, as well as specific demonstrations of the drawbacks (discussed in the previous section) that expanding over real exponential functions, which are not a Riesz basis, introduces into numerics.

The first of these examples is simple, but demonstrates an important point. In standard QSP, trivial protocols of length $n$ serve to generate the Chebyshev polynomials of the first kind $T_n$ in terms of the modified signal $x = \cos{\theta}$, by merit of their usual definition $T_n(x) = \cos{(n\arccos{x})}$. Such trivial protocols in the case of phased boosts are modified according to $x \mapsto ix$, giving
    \begin{equation}
        W(x) \mapsto V(x) =
        \begin{bmatrix}
            x & \sqrt{x^2 - 1}\\
            \sqrt{x^2 - 1} & x
        \end{bmatrix},
    \end{equation}
where it should be noted that $x$ here has been overloaded to refer to $\cosh{\beta}$, and thus has a different domain and codomain. In this setting the trivial protocol of length $n$, i.e., with $\Phi \in \mathbb{R}^{n + 1} = \{0, 0, \cdots, 0\}$, takes the form
    \begin{equation}
        V(x)^n = 
        \begin{bmatrix}
            T_n(x) & U_{n - 1}(x)\sqrt{x^2 - 1}\\
            U_{n - 1}(x)\sqrt{x^2 - 1} & T_n(x)
        \end{bmatrix},
    \end{equation}
where $x \in [1, \infty)$ now constitutes the valid range, and the $n$-th Chebyshev polynomial of the second kind is notated $U_n(x)$. Unlike their action on the interval $[-1, 1]$, the Chebyshev polynomials outside this interval have the following unique property.

\begin{theorem}[Extremal growth of Chebyshev polynomials]
    Among polynomials $P(x)$ of a fixed degree $n$ whose modulus obeys $|P(x)| \leq 1$ on $x \in [-1,1]$ the Chebyshev polynomial $T_n(x)$ is the unique polynomial (up to an overall phase) whose modulus increases most quickly on the complement of that interval, i.e., for $x \not\in [-1,1]$.
\end{theorem}

From this theorem we see that the trivial protocol in SU(1,1) QSP achieves, as might be expected, an extremal polynomial in a concrete sense: the magnitude of the top left element of the resulting transfer matrix increases as quickly as possible in $x$ for a given length protocol. Note that the restriction that this is over all such polynomials such that for $x \in [-1,1]$ they have bounded modulus is satisfied as $x \leq 1$ allows us to apply the inverse of the original $x \mapsto ix$ map, and recognize the resulting polynomial as that of standard QSP on the relevant interval, whose modulus bound follows from unitarity.

We can also look at another special function achievable by standard QSP protocols, extending this function to beyond $x \in [-1,1]$. Specifically, we look at the prescription for the phases of fixed-point amplitude amplification. In this setting, the QSP phases for a given protocol are defined recursively in terms of those of a shorter protocol, such that longer sequences better approximate one which, for nearly all $x$, generates a (possibly phased) bit-flip operation. In what follows, we translate the so-called $\pi/3$-protocol \cite{yoder_14, grover_05} directly to the setting of QSP. We note that \cite{yoder_14} gives an improved version of this protocol when one does not desire monotonicity of the success probability, but for our purposes, the achieved function has a much neater form. Moreover, the explicit QSP protocol we provide below is not described explicitly in related work.

\begin{definition}[Monotonically amplifying QSP protocol] \label{def:monotonic_amplification}
    Consider the following QSP phase list, which will be the base case of our recursion
        \begin{equation}
            \Phi_0 = \{0, -\pi/6 + \pi/2, \pi/6 - \pi/2, 0\}.
        \end{equation}
    Moreover, toward definition of the recursive step, given a phase list $\Phi$ consider the inverse phase list $\Phi^{-1}$ to be the reversed, negated version of $\Phi$ with its first and last elements modified in the following way
        \begin{align}
            \Phi &= \{\phi_0, \phi_1, \cdots, \phi_{n-1}, \phi_{n}\}\\
            \Phi^{-1} &= \{-\phi_{n} + \pi/2, -\phi_{n-1}, \cdots, -\phi_1, -\phi_0 -\pi/2\}.
        \end{align}
    Additionally, given two phase lists $\Phi_0, \Phi_1$ we define their concatenation as protocols, denoted $\Phi_0 \cup \Phi_1$, by the following operation, which unites the trailing phase of the first and leading phase of the second phase list:
        \begin{align}
            \Phi_0 &= \{\phi_{0,0}, \phi_{0,1}, \cdots, \phi_{0,n-1}, \phi_{0,n}\}\\
            \Phi_1 &= \{\phi_{1,0}, \phi_{1,1}, \cdots, \phi_{1,n-1}, \phi_{1,n}\}\\
            \Phi_0 \cup \Phi_1 &= \{\phi_{0,0}, \phi_{0,1}, \cdots, \phi_{0,n-1}, \phi_{0,n} + \phi_{1,0}, \phi_{1,1}, \cdots, \phi_{1,n-1}, \phi_{1,n}\}.
        \end{align}
    We can now define the recursive step which, given the phases for a monotonically amplifying QSP protocol $\Phi_n$, generates the phases of a longer monotonically amplifying protocol $\Phi_{n + 1}$:
        \begin{equation}
            \Phi_{n + 1} = \Phi_{n} \cup \{-\pi/6\} \cup \Phi_{n}^{-1} \cup \{\pi/6\} \cup \Phi_n.
        \end{equation}
    For clarity we list the initial phase list below, followed by the result after applying this recursive procedure once and then twice:
        \begin{align}
            \Phi_{0} &= \{0, 0\}\\
            \Phi_{1} &= \{0, 0, -\pi/6, \pi/2, -\pi/2, \pi/6, 0, 0\}\\
            \Phi_{2} &= \{0, 0, -\pi/6, \pi/2, -\pi/2, \pi/6, 0, 0, -\pi/6, \pi/2, 0, -\pi/6,\nonumber\\
            &\hphantom{{}={}\{} \pi/2, -\pi/2, \pi/6, 0,-\pi/2, \pi/6, 0, 0, -\pi/6, \pi/2, -\pi/2, \pi/6, 0, 0\}
        \end{align}
    It is evident that the length of this protocol increases exponentially in $n$. In the following lemma we give the functional transform this protocol achieves, and describe why it is termed monotonically amplifying.
\end{definition}

\begin{lemma}[Functional form of monotonically amplifying QSP protocol (Def.~\ref{def:monotonic_amplification})] \label{lemma:monotonic_amplification}
    The protocol with phases $\Phi_n$ as given in Def.~\ref{def:monotonic_amplification} generates a unitary with the following form
        \begin{equation}
            U_{\Phi_n} = 
            \begin{bmatrix}
                P_n & \;\cdot\; \\
                \;\cdot\; & \;\cdot\;
            \end{bmatrix},
        \end{equation}
    where the magnitude-squared of $P$ has the following simple form
        \begin{equation}
            |P_n(x)|^2 = x^{2(3^{n + 1})}.
        \end{equation}
    It is easy to check that this polynomial satisfies the required boundedness and parity conditions on $x \in [-1, 1]$. Moreover, for larger $n$, we see that every point on this graph in the range $x \in [-1,1]$ monotonically approaches $0$, meaning this matrix monotonically approaches a (possibly phased) bit flip as expected.
\end{lemma}

The function described in Lemma~\ref{lemma:monotonic_amplification} is very simple, and has some nice properties. When we consider this function under the usual map $\cos{x} \mapsto \cosh{\beta}$, meaning we evaluate it outside of the interval $[-1,1]$. While the modulus of $P$ necessarily does not grow as quickly outside $[-1,1]$ as the Chebyshev polynomial of the same degree, it has exceedingly regular form in terms of $x = \cosh{\beta}$, lending itself to easier analytic treatment.

Finally, we consider a simple phase sequence whose induced function has interesting and tunable (non-monotonic) behavior outside the interval $[-1,1]$. Specifically, take the QSP protocol of length $n + 1$ whose phases are identical:
    \begin{equation}
        \Phi = \{\phi, \phi, \cdots, \phi\} \in \mathbb{R}^{n + 1}.
    \end{equation}
For certain phases (notably $\phi \in \{0, \pi/2\}$), this reduces to simple and known protocols. For our purposes, however, we are interested in the behavior of the induced polynomial transforms for all possible choices of $\phi$ (though by symmetry we can restrict $\phi \in [0, \pi/2]$). It is not hard to determine that the basic iterate of this protocol, $V_\phi = e^{i\phi \sigma_z}W(x)$, has the following eigenvalues:
    \begin{equation}
        \lambda_{\pm} = \cos{\phi}
        \left(x \pm \sqrt{x^2 - \sec^2{\phi}}\right).
    \end{equation}
Moreover, one can quickly determine that the magnitude of these eigenvalues is one precisely when $|x| \leq |\sec{\phi}|$. In continuing the induced polynomial transform $P(x) = \langle 0 | U_\Phi | 0\rangle$ to all real $x$, one can see that there exists an \emph{extended region} in which the eigenvalues of the iterate have magnitude one. We now show that this induces a set of three regions in which the induced polynomial transform has substantively different character. We phrase this property in terms of the following three inequalities:
    \begin{alignat}{2}
        \sqrt{B(\phi, x)}
        \leq &|P(x)| 
        \leq 1, \quad &&0 \leq |x| \leq 1, \forall n,\\
        1 \leq &|P(x)| \leq \sqrt{B(\phi, x)}, \quad &&1 \leq |x| < \sec{\phi}, \forall n,\\
        \sqrt{T_n(x\cos{\phi})} \leq &|P(x)| < \infty, \quad &&\sec{\phi} \leq |x| < \infty , \forall n,
    \end{alignat}
where $T_n(x)$ is the $n$-th Chebyshev polynomial evaluated at $x$ and $B(x)$ is the following scaled and shifted secant function, which should be noted is independent of $n$:
    \begin{equation} \label{eq:secant_bound}
        B(\phi, x) \equiv 
        \frac{\sec{([\pi/2]\, x \cos\phi)} - 1}{\sec{([\pi/2]\cos\phi)} - 1}.
    \end{equation}
For a derivation of a simpler version of this upper bound, we refer the reader to the Appendix~\ref{appendix:upper_bounds}; however most of these bounds are simply recoverable using standard computer algebra software, taking care with the regions in which $P(x)$ is evaluated. Consequently we see that for $|x| \leq |\sec{\phi}|$, the maximum modulus attained by $|P(x)|$ is bounded from above by an expression that is independent of $n$, while for $x$ outside of this region, this modulus is bounded below by a series of functions which, for any fixed $x$, grow exponentially in $n$. Consequently, with a different character than that of the functional transforms seen in standard QSP, we have thresholding behavior in $x$ beyond the interval $[-1,1]$; that is, given any real number $\mu > 1$ and promised gap $\delta > 0$, there exists a positive integer $n$ and real angle $\phi$ such that the SU(1,1) QSP protocol with repeating phase $\phi$ and length $n$ produces a $P(x)$ whose magnitude at arguments $x_{\pm} = \mu \pm \delta$ differs by \emph{at least} any desired finite amount. We capture this statement in the following definition and theorem, and then discuss the character of the induced polynomial function more closely.

\begin{definition}[Weak step function] \label{def:weak_step_function}
    Given a piecewise-continuous function $f$ across two (possibly infinite) intervals $A, B \in \mathbb{R}$, where all elements of $A$ are less than all elements of $B$, we say $f$ is a $(g, h)$-weak step function if the following inequalities hold:
        \begin{equation}
            f_A \leq g_A, \quad f_B \geq h_B.
        \end{equation}
    In other words, the function $f$ is bounded from above by $g$ on the interval $A$, and bounded from below by $h$ on the interval $B$. For example, the Heaviside step function $\Theta(x)$ is a $(0, 1)$-weak step function for $A = \mathbb{R}_{-}$ and $B = \mathbb{R}_{+}$.
\end{definition}

In general, a function satisfying the properties of a weak step function may not be so interesting, especially if the characters of $g, h$ as given above are not substantively different. By the previous results, however, we will see that the properties of the weak step function induced by the QSP protocol of length $n + 1$ with constant phase factors are dramatic. Specifically, we showed above the following

\begin{lemma}[Weak step functions in constant-phase QSP]
    Let $P$ be the top-left element of the unitary matrix generated by the QSP protocol with constant phase list $\{\phi, \phi, \cdots, \phi\}$ of length $n + 1$. Then $|P|$ for $A = [1, \sec{\phi})$ and $B = [\sec{\phi}, \infty)$ is a $(B(\phi, x), T_n(\cos{\phi}))$-weak step function. Here we have made reference to Def.~\ref{def:weak_step_function} and the secant bound in Eq.~\ref{eq:secant_bound}.
\end{lemma}

Before showing a quantitative theorem about constant-phase QSP, we take a moment to analyze the bounds we've already given on the induced $|P|$ more closely, so that we might simplify our proofs. First, we note that most of the interesting behavior of $|P|$ occurs near the critical point $x = \sec{\phi}$, where we transition from our known upper bound (for all $n$) to our known lower bound (for each $n$). Taking $x = \sec{\phi} - \epsilon$, the upper bound provided approaches the following
    \begin{equation} \label{eq:negative_deviation}
        |P(\sec{\phi} - \epsilon)|^2 \geq B(\phi, x) =  
        \epsilon^{-1}\left[\frac{(2/\pi)\sec{\phi}}{\sec{(\pi\cos{\phi}/2)} - 1}\right] - 
        \left[\frac{1}{\sec{(\pi\cos{\phi}/2)} - 1}\right] + 
        \mathcal{O}(\epsilon).
    \end{equation}
We note that this is expected, as secant looks like the inverse function near its singularity. On the other side of this critical point, we can analyze the computed $|P|$ exactly, rather than its bound, and take an expansion
    \begin{equation} \label{eq:positive_deviation}
        |P(\sec{\phi} + \epsilon)|^2 = 
        \sqrt{1 + n^2 \tan^2{\phi}} + \mathcal{O}(\epsilon),
    \end{equation}
where we have kept only the zeroth order term for brevity. These particular limits can be computed laboriously by hand, or by computer algebra software. The key takeaway from both of these limits is that the behavior, for a fixed $\epsilon$ on either side of the critical point $x = \sec{\phi}$, of the function $|P|$ induced by the constant-phase QSP protocol is alternately bounded from above for $x < \sec{\phi}$ by a function constant in $n$, and bounded from below for $x > \sec{\phi}$ by a growing function in $n$. That the latter lower bound increases without bound in $n$ necessitates that the former upper bound approaches infinity near the critical point. The utility of the weak-step function induced here relies on that upper bound growing quite quickly.

Before finally posing a quantitative theorem, we also look at the limiting behavior of the functions considered here, and pair them with a few diagrams. In addition to expanding about the critical point, it is worthwhile to look at the behavior of $|P|$ for large $x$. In this case, we can replace terms of the form $\sqrt{x^2 - \sec{\phi}^2}$ in the analytical expression for $P$ by approximately $|x|$, and simplify the resulting expression. In this case one finds that, for $x$ larger than the critical point, the expression $|P|$ for the QSP protocol of length $n$ with constant phases $\phi$ approaches the clean function
    \begin{equation} \label{eq:large_x_limit}
        \lim_{x \gg \sec{\phi}} |P| = \frac{1}{2^{n + 1}} x^{n} \cos^{n + 1}{\phi}.
    \end{equation}
That is, rather than simply growing quadratically in $n$ as we observed just above the critical point, when one evaluates $P$ at some constant distance above the critical point, its magnitude grows exponentially in $n$ for all such arguments $x$. Indeed, as given by our Chebyshev-dependent lower bound, we see that this is within a constant factor dependent on $\phi$, as fast as any function achievable with QSP could grow on this interval.

\begin{theorem}[Properties of constant-phase SU(1,1) QSP] \label{thm:weak_step_function}
    Take $\xi > 0$ the desired minimal step distance and $\mu > 1$ the desired step location. For all $\delta > 0$, there exists a positive integer $N$ such that for all $n > N$, there is a QSP protocol of length $n$ which satisfies the following properties:
        \begin{align}
            |P(x < \mu - \delta)| &\leq \sqrt{B(\phi, x)}\\
            |P(x > \mu + \delta)| - |P(\mu - \delta)| &\geq \xi.
        \end{align}
    Moreover, the minimum length of this protocol goes as $N = \mathcal{O}(\mu\xi/\delta)$. Here $P(x)$ is the top-left element of the SU(1,1) QSP protocol with phase list $\{\phi, \phi, \cdots, \phi\}$ of length $n + 1$ with $\phi = \sec^{-1}{\mu} \in (0, \pi/2)$. Finally, an upper bound the required $N$ goes as
        \begin{equation}
            N \leq 
            \mathcal{O}\left[\delta^{-1/2}\sqrt{\xi \cot{\phi} + \frac{\csc{\phi}}{\sec{([\pi/2]\cos{\phi})} - 1}}\right].
        \end{equation}
    Here again we can restrict $\phi \in (0, \pi/2)$ by symmetry arguments. For simplicity we have also assumed that $\phi$ is not tending extremely close to $0$ or $\pi/2$, in which case the limiting behavior of the protocol becomes more involved. In this sense there exists suppressed constants dependent on the closeness of $\phi$ to these critical values; moreover at the critical points, as discussed before, the considered regions become degenerate. Moreover, it is easy to note that for limits $\phi \rightarrow \{0, \pi/2\}$, the upper bound for $N$ given grows arbitrarily large.
    
    Proof follows by working in the limiting region of small $\delta$, in which case the behavior on either side of the critical point of $|P|$ have analytical expressions (Eqs.~\ref{eq:negative_deviation} and \ref{eq:positive_deviation}). Requiring that these points be separated by $\xi$ immediately yields the given scaling, and thus an upper bound for $n$, as the rate of growth of $|P|$ beyond $x = \sec{\phi}$ is near maximal, as given in Eq.~\ref{eq:large_x_limit}. It is worthwhile to note that outside this critcial region, and especially for $\delta = \mathcal{O}(\mu)$, the required $N$ can grow as slowly as logarithmically in $\xi$.
\end{theorem}

\begin{figure}
    \centering
    \includegraphics[width=\textwidth]{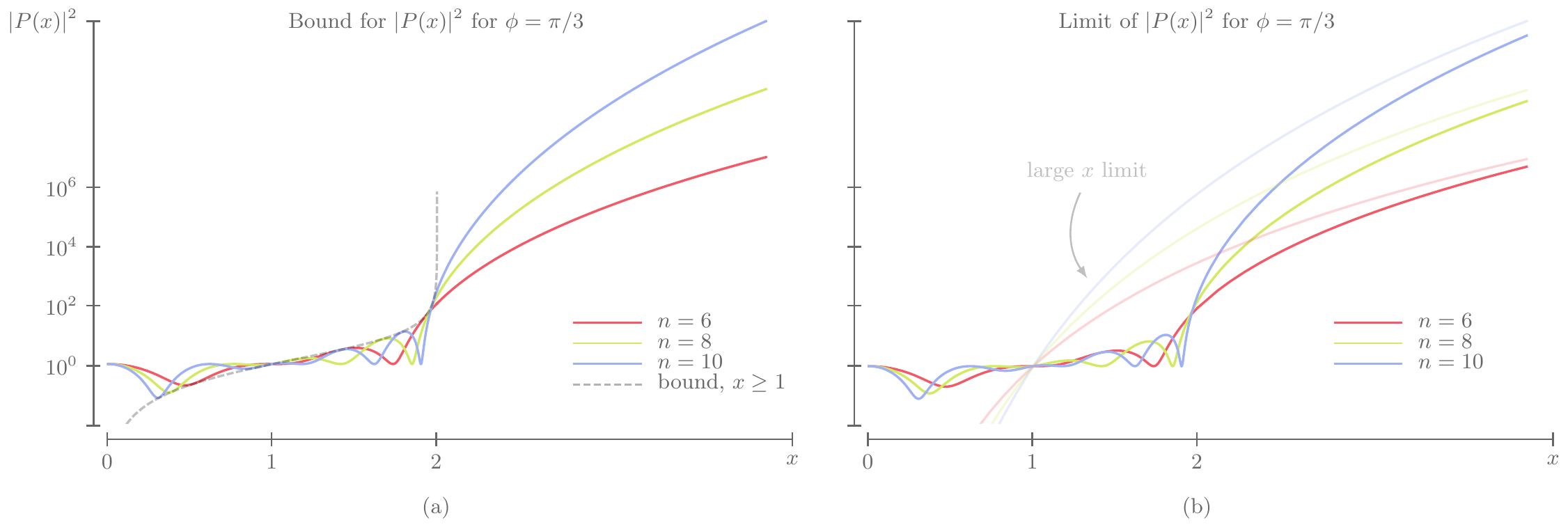}
    \caption{Two log plots of $|P|^2$ for QSP protocols with phase list $\{\pi/3, \pi/3, \cdots, \pi/3\}$ of lengths $6, 8, 10$, along with the analytically derived (a) upper bound for $|P|^2$ (Eq.~\ref{eq:secant_bound}), and (b) large-$x$ limit of these functions (Eq.~\ref{eq:large_x_limit}). Note that, for each $x$ along these plots above $x = 1$, the evaluated function grows exponentially in $n$, and that $|P|^2$ only converges to these limits well-after the critical point. One can also see the changed behavior at $x = \sec{\pi/3} = 2$, when the plotted functions become monotonic.}
    \label{fig:upper_bound}
\end{figure}

\begin{remark}[On weak step functions]
    Note that unlike we might hope to be the case, the achieved function in Theorem~\ref{thm:weak_step_function} does not uniformly converge to a discontinuous jump. Instead, we are given that, across to regions (here $x \leq \sec{\phi}$ and $x \geq \sec{\phi}$), that the magnitude of the relevant matrix element is respectively upper bounded and lower bounded by known and simple analytic functions (Def.~\ref{def:weak_step_function}). Moreover, these bounds are useful because the former is constant in $n$, while the later is shown to grow without bound in $n$ at a reasonable rate. Consequently, while this form of thresholding is not as strong as uniform approximation, it nevertheless captures a useful, tunable property of a polynomial function extended beyond its usually considered region in QSP.
\end{remark}

We pause to note that it is interesting that such a simple protocol, which to the authors' knowledge has had no application in standard QSP, nevertheless reveals novel properties under extension arguments of large modulus. It is reasonable to assume that other repeated QSP-like units could have, when analyzed in a similar way, experimentally useful properties when the argument is extended beyond the compact interval usually considered in QSP.

\section{Concluding discussion} \label{sec:discussion_conclusion}

In this work we have investigated a variant of the QSP ansatz where the interleaved components are generated not by complex exponentiation of elements of the $\mathfrak{su}(2)$ Lie algebra, but rather $\mathfrak{su}(1,1)$. Quantum systems which evolve within SU(1,1) occur frequently in photonic and mechanical settings, and while there necessarily do not exist finite dimensional unitary representations of such evolutions, one can identify the Heisenberg picture evolution of optical modes through series of parametric amplifiers with our proposed ansatz. One finds that while the achievable polynomial coefficients are identical between the SU(2) and SU(1,1) ansätze, a key difference emerges: said polynomials are evaluated not on the image of the cosine function, but instead that of the hyperbolic cosine. This simple equivalence can be entirely captured by the following relations
    \begin{align}
        \begin{bmatrix}
            P(\cos{\theta}) & iQ(\cos{\theta})\sin{\theta}\\
            iQ^*(\cos{\theta})\sin{\theta} & P*(\cos{\theta})
        \end{bmatrix}
        &\Longleftrightarrow 
        \begin{bmatrix}
            P(\cosh{\beta}) & Q(\cosh{\beta})\sinh{\beta}\\
            Q^*(\cosh{\beta})\sinh{\beta} & P*(\cosh{\beta})
        \end{bmatrix},\\[0.8em]
        \begin{bmatrix}
            P(x) & iQ(x)\sqrt{1 - x^2}\\
            iQ^*(x)\sqrt{1 - x^2} & P*(x)
        \end{bmatrix}
        &\Longleftrightarrow 
        \begin{bmatrix}
            P(x) & Q(x)\sqrt{x^2 - 1}\\
            Q^*(x)\sqrt{x^2 - 1} & P*(x)
        \end{bmatrix},\\[1em]
        \Re[P]^2 + (1 - x^2)\Re[Q]^2 \leq 1 &\Longleftrightarrow \Re[P]^2 + (x^2 - 1)\Re[Q]^2 \geq 1\\[1em]
        x \in [-1,1] 
        &\Longleftrightarrow  
        x \in [1, \infty).
    \end{align}
We note that even this apparently simple transformation (of only the argument of the achieved function, not the coefficients) nevertheless presents significant barriers to understanding the nature (especially in the approximation of desired functions) of these polynomials. Standard QSP can care only, by definition, about $x \in [-1,1]$, and it is not evident (nor even true) that arbitrary functional approximation outside this interval can be achieved, and moreover achieved efficiently (i.e., with short protocols). The business of re-proving similar statements to those in QSP relies on general tools in functional analysis. Ultimately this apparently simple transformation reduces the problem of characterizing the achievable functions of QSP to the study of polynomials over the reals which are (1) bounded above in magnitude by one on the interval $[-1,1]$, (2) bounded below in magnitude by one outside the interval $[-1, 1]$, and (3) of definite parity. We show that the modified ansatz is still dense in the set of continuous functions also obeying constraints (1-3), although the length of the achieving protocols can, in the worst case, grow very fast. This poor scaling is due to the desire to approximate functions over non-compact sets.

Toward useful application of QSP-like ansätze to continuous-variable computations, we provided a series of concrete phase prescriptions for which the achieved polynomial transforms are analytically simple: these include Chebyshev polynomials, monomials, and a thresholding function whose step location and height are precisely tunable. The last of these is noteworthy, as its standard QSP counterpart had no obvious use previously, indicating that simple constructions in standard QSP can reveal unexpected traits when extended to regions in parameter space the standard ansatz had no way of accessing. These protocols, which are shown, unlike their SU(2) counterparts, to be especially sensitive to their underlying parameters, thus form a natural building block for general amplification techniques in quantum systems.

Specifically, recent advances not only in optics but also superconducting quantum computing systems have investigated the possibility of cascaded series of parametric amplifications for quantum measurement, most successfully in the form of travelling wave parametric amplifiers (TWPAs) \cite{mohdbzos_twpa_15, peng_floquet_twpa_22}. While we consider a simpler model, the ability to precise tune the location and magnitude of thresholding behavior resulting from a cascaded series of parametric amplifiers, as shown in Thm.~\ref{thm:weak_step_function}, holds promise for constituting the analytic language of such devices. Simple and exciting possibilities for extensions include the development of band-pass functional transforms \cite{wimperis_bb1_94}, protocols for signal trifurcation, and application of chaotic theory to our protocols in the long-length limit.

On this line, going forward there is great promise in better understanding functional analysis over non-compact sets and with non-Riesz bases, toward a coherent theory of alternating ansätze over families of physically relevant Lie algebras. Understanding how one can precisely control such quantum systems opens the door for the development of new algorithms in the continuous variable setting, while maintaining the single-qubit intuition that has made quantum signal processing so successful.

\section{Acknowledgements}

ZMR was supported in part by the NSF EPiQC program, and ILC was supported in part by the U.S. DoE, Office of Science, National Quantum Information Science Research Centers, and Co-design Center for Quantum Advantage (C2QA) under Contract No.\ DE-SC0012704. WJM and VMB acknowledge partial support through the MEXT Quantum Leap Flagship Program (MEXT Q-LEAP) under Grant No.\ JPMXS0118069605. The authors would also like to thank NTT Research Inc.\ and NTT BRL Atsugi for their support in this collaboration, and specifically comments and suggestions from Sho Sugiura and Sina Zeytinoğlu.

\bibliography{main}

\appendix

\section{Basic results in QSP and QSVT} \label{appendix:basics_qsp_qsvt}

In the following section we provide proofs of a few main theorems in the body of the paper, as well as additional commentary on relevant mechanisms of the standard proof methods of QSP and QSVT. We aim for these to be self-contained, and to give some intuition for how one attempting to expand or modify the theory of these circuits might begin to do so.

    \begin{remark} \label{remark:qsp_ansatz_expressivity}
        \begin{proof}
            In reference to Theorem~\ref{thm:qsp_ansatz_expressivity}. The unitary matrix corresponding to a QSP protocol has the form
                \begin{equation}
                    U_{\Phi} \equiv
                    \begin{bmatrix}
                        P(x) & \sqrt{1 - x^2}Q(x) \\
                        -\sqrt{1 - x^2}Q^*(x) & P^*(x)
                    \end{bmatrix},
                \end{equation}
            where $P, Q \in \mathbb{C}[x]$ have definite parity, are bounded above in magnitude by $1$, and obey the relation $|P|^2 + (1 - x^2)|Q|^2 = 1$. It is also known \cite{gslw_19} that choosing the real part $\tilde{P}$ of $P$ with definite parity and bounded above by $1$ defines a QSP protocol (i.e., a finite list of real phases $\Phi \in \mathbb{R}^{n + 1}$) whose unitary has the following form
                \begin{equation}
                    U_{\Phi} \equiv
                    \begin{bmatrix}
                        \tilde{P}(x) + iB(x) & i\sqrt{1 - x^2}C(x) \\
                        i\sqrt{1 - x^2}C(x) & P(x) - iB(x)
                    \end{bmatrix},
                \end{equation}
            where $\tilde{P}, B, C \in \mathbb{R}[x]$ and $\Re[P] = \tilde{P}$ as stated. The map $\tilde{P} \rightarrow \Phi$ is not unique, but can be made unique by choosing how $B, C$ are defined in terms of the zeros of $1 - \tilde{P}^2$. It is not difficult to see that this unitary is a rotation of the form
                \begin{align}
                    U_\Phi &= \tilde{P}(x) I + i\, \big[B(x) Z + \sqrt{1 - x^2} C(x) X\big]\\
                           &= \cos{(\xi(x))} I + i\sin{(\xi(x))}\,e^{iR(x)Y}Ze^{-iR(x)Y},
                \end{align}
            where we have defined new functions of $x$, namely $\xi(x)$ and $R(x)$ with the following form
                \begin{align}
                    \xi(x) &= \arccos{\biggr[\tilde{P}(x)\biggr]},\\
                    R(x) &= \arccos{\left[B(x)/\sqrt{1 - \tilde{P}^2}\right]}.
                \end{align}
            Here we identify $R(x)$ as the overall $Y$-rotation ambiguity for the unitary, parameterized by $x$ in a way that completely depends on both the choice of $\tilde{P}$ and certain choices concerning the roots of $1 - \tilde{P}^2$ as stated to make $\Phi$ unique. Both $\xi(x)$ and $R(x)$ are in truth functions $[-1, 1] \rightarrow U(1)$, choosing an angle on the circle through the branch cut of the arccosine function. The speed of convergence to a desired real function by $\tilde{P}$ is a well known classical result in functional analysis.
        \end{proof}
    \end{remark}

    \begin{remark} \label{remark:su11_qsp_completion}
        \begin{proof}
            In reference to Theorem~\ref{thm:matrix_completion_su11}. The intent of this theorem, in analogy to that of standard QSP, is to determine if partial constraints on the representation of SU(1,1) as given allow one to complete unspecified elements of the matrix such that the overall matrix can be realized as a product of phased boosts. We assume the existence of a \emph{real} polynomials $P$ and $Q$ of definite parity (though opposite from one another), such that
                \begin{equation}
                    P^2 - (x^2 - 1)Q^2 \geq 1, x \in [1, \infty).
                \end{equation}
            As in standard QSP, we can examine the polynomial function $F$ with the following form
                \begin{equation}
                    F = P^2 - (x^2 - 1)Q^2 - 1,
                \end{equation}
            which is necessarily positive semidefinite on the relevant interval by our assumption. By a simplified version of the Fejér-Riesz theorem \cite{rc_22} or simple root analysis \cite{gslw_19}, there exists a complex polynomial $G(x)$ of definite parity such that the function $F$ is the magnitude square of $G$, i.e.,
                \begin{equation}
                    F = G(x)G^{*}(x).
                \end{equation}
            Note that the existence of such a decomposition depends solely on the positive semidefiniteness of the relevant polynomial $F$. The imginary parts of $\tilde{P}, \tilde{Q}$ in the theorem statement can now be identified with the real and imaginary parts of the polynomial $G$, the latter of which will permit $\sqrt{x^2 - 1}$ to be factored from it by merit of the known boundary conditions at $x = 1$, equivalently $\beta = 0$. That the resulting matrix corresponds to a product of phase boosts now follows directly from the mapping from $\cosh{\beta}$ to $\cos{\theta}$, under which the coefficients do not change, but we are guaranteed that the modified polynomials must now satisfy the standard $P^2 + (1 - x^2)Q^2 \leq 1$ on $x \in [-1,1]$. 
            
            In this way we see that the constraints for matrix completion in SU(2) and SU(1,1) QSP are dual to one another. But while we have a succinct description of the real $P, Q$ that are achievable, the required degree for uniform approximation by such functions of a desired continuous function of the same parity/bound constraints, is unclear, and left to Sec.~\ref{sec:non_harmonic_analysis}. 
        \end{proof}
    \end{remark}

\section{Upper bounds in concrete SU(1,1) QSP protocols} \label{appendix:upper_bounds}

For the QSP protocol of length $n$ with fixed phases $\phi$, the text gives a particularly simple upper bound for $|P(x)|$ (for $1 \leq |x| < \sec{\phi}$) which also suffices as a lower bound for $|x| \leq 1$. The derivation of this bound, however, is both involved and not particularly illuminating, and consequently we give a more simply derived upper bound for the relevant region $1 \leq |x| < \sec{\phi}$ whose properties are nevertheless good enough to prove all results in the text on the scaling of the required length $n$ in terms of relevant properties of the achieved functions. The first step in this derivation is determining the general form of $P$ of the generated unitary for fixed $\phi$; this corresponds to exponentiating the basic iterate $e^{i\phi\sigma_z}V(x)$ with an additional overall phase, which can be done via computer algebra software or by hand by diagonalizing the relevant small matrix. This element has the form
    \begin{align}
        P(x) {}={}
        \frac{1}{2}e^{-i(n-1)\phi}&\frac{1}{\sqrt{x^2 - \sec^2\phi}}\nonumber\\
        \biggr[
            &\left(\sqrt{x^2\cos^2\phi - 1} - ix\sin{\phi}\right)\left(x\cos\phi - \sqrt{x^2\cos^2\phi - 1}\right)^n + \nonumber\\
            &\left(\sqrt{x^2\cos^2\phi - 1} + ix\sin{\phi}\right)\left(x\cos\phi + \sqrt{x^2\cos^2\phi - 1}\right)^n
        \biggr].
    \end{align}
While this expression does indeed have dependence on $n$, we can now show that this dependence is mild, and does not affect the magnitude of this term very strongly in the critical region $1 \leq |x| < \sec{\phi}$. This follows by noting that the terms raised to the $n$-th power in the second and third line are, for $1 \leq |x| < \sec{\phi}$, bounded in magnitude by a constant, namely $1$. Replacing these terms with this upper bound, and repeatedly applying the simple triangle-inequality-derived upper bound for the magnitude of a complex number $|a + bi| \leq a + b$, we find that this whole expression is upper bounded by the following
    \begin{align}
        |P(x)| &\leq \frac{1}{\sqrt{\sec^2\phi - x^2}} \left[\sqrt{\sec^2\phi - x^2} + x\tan\phi\right],\\
        &= 1 + \frac{x\tan\phi}{\sqrt{\sec^2\phi - x^2}} \label{eq:simple_bound},
    \end{align}
which has the expected behavior at $x = 0$, as well as the expected singularity at $x = \sec{\phi}$. Comparing this bound to the one given in the body of the paper (Eq.~\ref{eq:secant_bound}), we see that their limiting behavior around $x = \sec{\phi}$ are characterized by the following limits
    \begin{align}
        |P(\sec{\phi} - \delta)| &\approx \frac{1}{\sqrt{2}} \sec^{1/2}\phi \tan\phi \frac{1}{\delta^{1/2}},\\
        &\approx \sqrt{\frac{2}{\pi}}\sqrt{\frac{\sec{\phi}}{\sec{[(\pi/2)\cos\phi]} - 1}}\frac{1}{\delta^{1/2}},
    \end{align}
where the first has been derived from Eq.~\ref{eq:secant_bound}, and the second from Eq.~\ref{eq:simple_bound} taking $x = \sec\phi - \delta$, and computing the leading order singular term in $\delta$. Comparing these two terms numerically one can find that they agree exceptionally well as $\phi$ approaches $\pi/2$, despite radically different forms. Their distinguishing features lie in the difference between uncommon terms multiplying these two functions:
    \begin{equation}
        \left|\tan\phi - 2\sqrt{\frac{1}{\pi\sec{[(\pi/2)\cos\phi] - 1}}}\right|,
    \end{equation}
where the first term is clearly unbounded while the latter term is bounded in magnitude by $2/\sqrt{\pi - 1}$ as $\phi \mapsto \pi/2$. The similarity of these multiplying terms, however, lies in the fact that both act approximately as $\phi$ in the limit as $\phi \rightarrow 0$. Consequently as long as we do not consider $\phi$ too close to $\pi/2$, which is an unphysical region for a variety of reasons, then our simply-derived upper bound will be close in its behavior around $x = \sec{\phi}$ to the more sophisticated bound.

\section{Representation theory for Lie groups and algebras} \label{appendix:basics_rep_theory}

In this section we cover a few basic definitions from representation theory, toward a minimal presentation of a few no-go theorems on the finite-dimensional unitary representation of certain non-compact Lie groups.

\begin{definition}[Group representation]
    A representation of a group $G$ on a finite-dimensional complex vector space $V$ is a homomorphism $\rho: G \rightarrow \text{GL}(V)$, from $G$ to the general linear group over $V$ (with respect to some field $K$). Usually we will just refer to the vector space $V$ as the representation of $G$, when we really intend $\rho$.
\end{definition}

\begin{definition}[Irreducible representation]
    A representation $V$ is called irreducible if there is no proper nonzero invariant subspace $W$ of $V$. Here the supposed $W$ would be a subrepresentation of $V$, namely a vector subspace $W$ of $V$ which is invariant under $G$, the group being represented, with respect to the defining homomorphism of the reprsentation.
\end{definition}

\begin{definition}[Unitary representation]
    A representation $V$ (over a complex Hilbert space) is said to be unitary if the image of the homomorphism $\rho$ is a unitary matrix for all $g \in G$. Such representations are often coveted in physics, given the physical interpretation natural to unitary operators.
\end{definition}

In what follows we reproduce a few prominent results from the last century on the representation theory of certain special Lie groups. The first, the Peter-Weyl theorem, concerns compact Lie groups like SU(2), and showcases a variety of the natural properties on which the success of QSP and QSVT can be seen to crucially rely. The second theorem, not explicitly named but often cited in the study of the Lorentz group, provides a prominent counterexample when some of the assumptions of the Peter-Weyl theorem (most notably compactness) are violated.

\begin{theorem}[Peter-Weyl theorem, from \cite{peter_weyl_01}]
   Take $\rho$ a finite dimensional continuous group representation of $G$. Then the Peter-Weyl theorem can be summarized in three statements concerning this representation
        \begin{enumerate}[label=(\arabic*)]
            \item (Density). The set of matrix coefficients of $G$ is dense in the space of continuous complex functions $C[G]$ on $G$, equipped with the uniform norm (this also implies density in $L^2[G]$.
            \item (Complete reducibility). Let $\rho$ be a unitary representation of a compact group $G$ on a complex Hilbert space $H$. Then $H$ splits into an orthogonal direct sum of irreducible finite-dimensional unitary representations of $G$.
            \item (Generalized Fourier basis). The matrix coefficients for $G$, suitably renormalized, are an orthonormal basis of $L^2[G]$.
        \end{enumerate}
    While the statements of the Peter-Weyl theorem are not in themselves strong enough to recover the theorems of QSP, due to the specificity of its alternating ansatz, they are nevertheless intimately connected to the reason that the achievable space of functions parameterizing QSP circuits is dense in the space of continuous functions with similar constraints.
\end{theorem}

In contrast to the assurances of the Peter-Weyl theorem, which guarantee most desired analogues to the requirements of harmonic analysis on compact groups, the case for non-compact groups is somewhat bleak. The following theorem justifies our focus on finite dimensional (non-unitary) representations of SU(1,1), as well as the difficulty in finding Riesz bases for natural spaces of functions over the elements of such representations.

\begin{theorem}[Theorem of \cite{wigner_39} (p 165)]
    Non-compactness implies, for a connected simple Lie group, that no nontrivial finite-dimensional unitary representations exist. In other words, unitary irreducible representations (except the identity) of non-compact Lie groups are infinite dimensional.
\end{theorem}

\end{document}